\documentclass[11pt, journal, a4paper]{IEEEtran}

\usepackage[absolute]{textpos}
\TPMargin{5pt}

\newcommand{\copyrightstatement}{
    \begin{textblock}{0.84}(0.08,0.93)    
         \noindent
         \footnotesize
          Copyright \copyright 2012 IEEE. Personal use of this material is permitted. However, permission to use this material for any other purposes must be obtained from the IEEE by sending an email to pubs-permissions@ieee.org
    \end{textblock}
}

\ifCLASSINFOpdf
  \usepackage[pdftex]{graphicx}
  \graphicspath{{.}, {./figures/}}
  \DeclareGraphicsExtensions{.pdf, .png}
\else
\fi

\usepackage[cmex10]{amsmath}
\usepackage{amssymb}
\usepackage[tight,footnotesize]{subfigure}
\usepackage{bm}
\usepackage{algorithm, algpseudocode, algorithmicx}

\newcommand{\mat}[1]{\ensuremath{\bm{\mathrm{#1}}}}
\newcommand{\A}{\ensuremath{\mat{A}}}
\newcommand{\B}{\ensuremath{\mat{B}}}
\newcommand{\x}{\ensuremath{\mat{x}}}
\newcommand{\X}{\ensuremath{\mat{X}}}

\newcommand{\e}{\ensuremath{\mat{e}}}

\newcommand{\y}{\ensuremath{\mat{y}}}
\newcommand{\Y}{\ensuremath{\mat{Y}}}

\newcommand{\Ps}{\ensuremath{\bm{\Psi}}}

\newcommand{\Ph}{\ensuremath{\mat{\Phi}}}

\newcommand{\trasp}[1]{\ensuremath{#1 ^\mathsf{T}}}

\newcommand{\vect}[1]{\ensuremath{\mathrm{vec}\left \{ #1\right\}}}

\newcommand{\N}{\mathcal{N}}

\def\df{\triangleq}
\def\Ri{\mathbb{R}}

\newcommand{\lzeronorm}[1]{\ensuremath{\left\| #1\right\|_{\ell_0}}}
\newcommand{\lonenorm}[1]{\ensuremath{\left\| #1\right\|_{\ell_1}}}
\newcommand{\ltwonorm}[1]{\ensuremath{\left\| #1\right\|_{\ell_2}}}

\newtheorem{remark}{Remark}[section]

\newcommand{\F}{\ensuremath{\mat{F}}}

\newcommand{\diag}[1]{\ensuremath{\mathrm{diag}\big ( #1 \big
)}}

\begin{document}
\copyrightstatement
%
\title{Progressive compressed sensing and reconstruction of multidimensional signals using hybrid transform/prediction sparsity model}
%
%
%

\author{Giulio~Coluccia,~\IEEEmembership{Member,~IEEE,}
		Simeon~Kamdem~Kuiteing,
        Andrea~Abrardo,~\IEEEmembership{Senior Member,~IEEE,}
        Mauro~Barni,~\IEEEmembership{Fellow,~IEEE,}
        and~Enrico~Magli,~\IEEEmembership{Senior Member,~IEEE}
\thanks{G. Coluccia and E. Magli are with Politecnico di Torino (Dipartimento di Elettronica e Telecomunicazioni), c.so Duca degli Abruzzi, 24 - 10129 Torino, Italy (email: giulio.coluccia@polito.it, enrico.magli@polito.it). Their research leading to these results has received funding from the European Research Council under the European Community's Seventh Framework Programme (FP7/2007-2013) / ERC Grant agreement n° 279848}
\thanks{S. Kamdem Kuiteing, A. Abrardo and M. Barni are with University of Siena (Dipartimento di Ingegneria dell'Informazione), via Roma, 56 - 53100 Siena, Italy (email: barni@dii.unisi.it, abrardo.dii.unisi.it,
 skamdemk@yahoo.fr).}}

\markboth{To be published in IEEE Journal on Emerging and Selected Topics in Circuits and Systems}%
{Coluccia \MakeLowercase{\textit{et al.}}: TITLE TITLE TITLE}
%



\maketitle

\begin{abstract}

Compressed sensing (CS) is an innovative technique allowing to represent signals through a small number of their linear projections. Hence, CS can be thought of as a natural candidate
for acquisition of multidimensional signals, as the amount of data acquired and processed by
conventional sensors could create problems in terms of computational complexity. In this paper we propose a framework  for the acquisition and reconstruction of multidimensional correlated signals. The approach is general and can be applied to $D$ dimensional signals, even if the algorithms we propose to practically implement such architectures apply to 2D and 3D signals. The proposed architectures employ iterative local signal
reconstruction based on a hybrid transform/prediction correlation
model, coupled with a proper initialization strategy.

\end{abstract}


%
\IEEEpeerreviewmaketitle

\section{Introduction}\label{sec:introduction}

Compressed Sensing (CS) \cite{candes2006cs, donoho2006cs} has recently emerged as an efficient technique for sampling a signal with fewer coefficients than the number dictated by classical Shannon/Nyquist theory. The assumption underlying this approach is that the signal to be sampled is sparse or at least ``compressible'', i.e., it must have a concise representation in a convenient basis. In CS, sampling is performed by taking a number of linear projections of the signal onto pseudorandom sequences. Therefore, the acquisition presents appealing properties such as low encoding complexity, since the basis in which the signal is sparse does not need to be computed, and universality, since the sensing is blind to the source distribution. Reconstruction of a signal from its projections can be done e.g. using linear programming \cite{donoho2006cs}, with a complexity that is $O(N^3)$, with $N$ the number of samples to be recovered.

Plenty of applications are possible, ranging from
image and video to biomedical and spectral imaging, just to mention
a few. A single-pixel camera has been demonstrated in
\cite{singlepixel, wakin2006aci}, which uses a single detector to sequentially
acquire random linear measurements of a scene. This kind of design
is very interesting for imaging at wavelength outside the visible
light, where manufacturing detectors is very expensive. CS could be
used to design cheaper sensors, or sensors providing better
resolution for an equal number of detectors. E.g., in \cite{brady}
an architecture is proposed based on Hadamard imaging, coupled with
reconstruction techniques borrowed from CS.

The acquisition of multidimensional signals could benefit from CS due to its low-complexity sampling process and the reduction of the number of samples to be taken, processed and transmitted. In this case, a serious problem arises regarding the computational complexity of the reconstruction process. The conventional approach of measuring the signal along all dimensions at once leads to very large $N$, making the reconstruction computationally intractable. The simplest solution is to take separate sets of measurements grouping subsets of dimensions and to perform separate reconstructions. For example, an hyperspectral image could be acquired in the spatial or spectral dimensions. However, this ``separate"
approach does not yield satisfactory performance in terms of
mean-squared error (MSE), as it neglects the overall correlation among every dimension. In the example above, the spatial CS approach completely
neglects the spectral correlation, and the spectral approach
neglects the spatial one.

The authors of \cite{duarte2010kronecker} showed a way to recast a multidimensional CS problem to a one-dimensional one, by the means of Kronecker products of sensing and sparsity matrices. The problem of this approach is that the dimensionality of the CS problem to be solved rapidly grows as the product of the sizes of each dimension. The authors of \cite{fowler2011multiscale} apply CS to blocks of images considering wavelets as sparsity basis.
Reconstruction algorithms for multidimensional signals have also been proposed in \cite{abrardo2011compressive,fowler_mv,wakin_mv} for hyperspectral images and multiview video.

In this paper, we propose a generic framework for CS acquisition and reconstruction of multidimensional signals. Then, we propose architectures for 2D and 3D signals. The framework is general since the principles behind the architectures we propose can be easily extended to $D$-dimensional signals, with arbitrary $D$. Moreover, the proposed architectures can be practically implemented following the algorithms we devise for each architecture. The architectures we propose can be applied to several scenarios we describe in the following paragraphs.

The first architecture describes a practical implementation for devices that acquire 2D visual information through progressive scanning \cite{coluccia2012novel}.
These devices are equipped with a one-dimensional array of detectors, and a 2D image is obtained via the repeated use of the array over different slices of the 2D object to be imaged. This is a very important scenario, which encompasses many applications. Amongst others, it is worth mentioning at least two examples, which we will focus on in the respective sections of this paper. The first is given by {\em flatbed scanners}, where each line of the image is acquired by a 1D optical sensor moving in the orthogonal direction. The second one is represented by {\em airborne and spaceborne imagers} of the pushbroom type for remote sensing applications. In this case, the 1D sensor is carried on a flying platform such as an airplane or satellite; the sensor looks down at the Earth, and acquires a line-by-line scan of the underlying scene, while each line is oriented in the across-track direction, and the platform flight moves the sensor from one line to the next one. These applications, as well as several other ones, can clearly benefit from CS. Devices similar in principle to the Single Pixel Camera can be applied to progressive scanning, where a 1D micromirror array can be used to directly sense lines in the CS format. In the case of the remote sensing imaging system, CS can lead to a simpler and cheaper system, which uses a single detector and produces a reduced number of sampling. Detectors can be costly in the wavelengths outside the visible spectrum, and the reduced number of samples allows to implement simpler onboard processing systems. For the flatbed scanner, CS would be extremely useful in order to develop a scanner of small size, as the CS sensor needs not be of the same physical size as the document being scanned. Moreover, in both cases, processing and data handling would be greatly reduced, which is important in order to reduce power consumption in the remote sensing case, and in order to enable application to small-sized low-power devices in the flatbed scanner case.

The second scenario is the acquisition of hyperspectral images. Satellite imaging is a highly effective tool in a variety of scientific and engineering contexts because of the information it provides about the nature of the materials being imaged. While traditional digital imaging techniques produce images with scalar values associated with each pixel location, in multi- and hyperspectral images these values are replaced with a vector containing the spectral information associated to that spatial location. The resulting image is therefore threedimensional
(two spatial and one spectral dimensions), and spectral resolution is very important for several applications,
including classification, anomaly detection, and spectral unmixing. Despite the huge potential, however, many modern
satellite imagers face a limiting trade-off between spatial and spectral resolution. In fact, the total number of samples that can be acquired is constrained by the size of the detector array. This limits the usefulness and cost-effectiveness of spectral imaging for many applications. This scenario intends to investigate the possibility of overcoming this limitation by means of a new imaging architecture based on CS, that is, an architecture in which the acquisition system does not detect single pixels of the scene, but rather a small number of measurements. Reconstruction of the image is going to be performed at the ground station, and all subsequent processing steps (radiometric and geometric calibration, orthorectification, and applications) would be performed on the reconstructed image. In this paper we consider several possible architectures that are based on the constraints imposed by real-world spectral imaging systems. A first architecture is derived from 2D imaging systems, and is based on the concept of a sensor that acquires the image as a whole, both spatially and spectrally. In this case, the single-pixel camera paradigm is still applicable, provided that this single pixel is actually a single-pixel spectral imaging device, i.e. one that separates the spectral components of the incoming light into several different wavelengths, in such a way that the integration of spatially modulated light can be done individually in each spectral channel, yielding a different set of linear measurements for each wavelength. The second architecture mimics spectral imagers of the pushbroom type. These scanners have a 2D sensor, but the two dimensions correspond to the spectral (wavelength) dimension and one spatial dimension, i.e. the direction perpendicular to the flight path. At a given flight position, the scanner acquires at once one spatial-spectral slice, i.e. one line of the image with all its spectral components; as the satellite or airplane moves one, another spatial-spectral slice is acquired, and this process is repeated until the desired image length has been obtained. In this paper we also consider a CS imaging architecture based on this paradigm. We assume that the single-pixel concept is applied to spectral slices, i.e. there is a diffractive element before the single pixel, so that the spatial-spectral slice can be modulated before being sensed by the single detector. It is worth noticing that these architectures would yield significant benefits to hyperspectral imaging systems. They would allow to reduce the cost and size of the sensor itself, for the same spatial and spectral resolution, thanks to the reduced number of detectors. Moreover, they would allow to greatly simplify onboard data handling, since the compressed acquisition process would generate much less data than a conventional system, allowing to employ less memory and computing power, eliminating the need of onboard compression altogether, and eventually leading to reduced power consumption, which is a critical aspect of any remote sensing mission. Despite the very appealing advantages, however, these acquisition architectures entail a reconstruction problem of huge size. Indeed, optimal reconstruction must exploit signal sparsity (and hence correlation) in all dimensions, requiring to solve the reconstruction problem at once employing a three-dimensional transform as sparsity model. This problem has huge computational complexity, and becomes infeasible for rather small image sizes, highlighting the need of techniques that can achieve near-optimal performance with a reasonable computational complexity.

In this paper we address these scenarios, and tackle the reconstruction problem for 2D and 3D signals. In particular, for 2D images, we propose a simple progressive acquisition algorithm, where rows are acquired independently of each other, but the reconstruction is performed jointly over all rows. Joint reconstruction is achieved through an iterative algorithm that correlates different rows through linear prediction filters, instead of taking a multidimensional transform as sparsity domain. Prediction filters allow to exploit correlation in both horizontal and vertical dimensions, even if the acquisition is performed in one direction only. The main concept is to exploit correlation along the vertical direction by iteratively predicting each line and reconstructing the prediction error only, which is more compressible than the line itself. Results show that few iterations of the proposed algorithm suffice to significantly improve the MSE of the reconstruction, allowing
to obtain high-quality reconstruction results with feasible complexity.
On the other hand, for 3D images we have more degrees of freedom than in the 2D case. For example, in \cite{kcs_icassp} it has been shown that 2D spatial CS (i.e., every
spectral channel is measured independently) has better performance
than spectral CS (in which every spectral vector is measured
independently), just because the former approach models correlation
in two dimensions, and the latter in only one. However, it should be
noted that even spatial CS achieves an MSE that is not small enough
for many hyperspectral applications, as the relative error is around
$\pm 5\%$ for sensible values of the number of acquired samples.
The key idea is that, in order to improve reconstruction quality,
correlation must be exploited in all three dimensions of the
spectral cube. To achieve this goal, we
propose several approaches, combining an accurate modelling of
the spatial-spectral correlations, with the low complexity of
sequential, as opposed to fully joint, reconstruction. In
particular, instead of modelling the correlation by means of a
three-dimensional transform, and hence attempting to reconstruct the
hyperspectral cube as a whole, we employ a linear correlation model
of the hyperspectral image, and iteratively apply this model band by
band, improving the quality of the reconstructed image. An alternative algorithm applies the model to the spectral rows of the image, iterating along rows.

Since the quality of the reconstructed signal depends on two factors: $i)$ the initialization of the iterative procedure and $ii)$ the accuracy of the linear prediction filters, we consider different initialization strategies based either on a 2D
 CS approach or on a simplified 3D strategy \cite{duarte2009kronecker} and test different prediction filters looking for the one providing better performance.

 This paper is organized as follows. Section~\ref{sec:background} contains the background of this work, notations and definitions. In section~\ref{sec:prop_alg} we describe in detail the algorithms we propose. In section~\ref{sec:num_res} we show some results obtained by simulations and we conclude our work in section~\ref{sec:conclusions}.

\section{Background}\label{sec:background}
\subsection{Notation and definitions}\label{sec:notation}

We denote (column-) vectors and matrices by lowercase and uppercase boldface
characters, respectively. The $(m,n)$-th element of a matrix $\A$ is $(\A)_{m,n}$. The $m$-th row of matrix $\A$ is $(\A)_m$. The $n$-th element of a vector $\mat{v}$ is $(\mat{v})_n$. The transpose of a matrix $\A$ is $\trasp{\A}$.

The stack operator $\vect{\A}$ denotes the column vector obtained by stacking the columns of \A\ on top of each other, from left to right.

We denote 3D variables by calligraphic letters, e.g. $\mathcal{Q}$. $\X = \mathcal{Q}_{i,:,:}$ is the matrix obtained by fixing index $i$ in the first dimension of $\mathcal{Q}$. $\Y = \mathcal{Q}_{:,j,:}$ is the matrix obtained by fixing index $j$ in the second dimension of $\mathcal{Q}$. $\mat{Z} = \mathcal{Q}_{:,:,k}$ is the matrix obtained by fixing index $k$ in the third dimension of $\mathcal{Q}$.

The notation $\lzeronorm{\mat{v}}$ denotes the number of nonzero elements of vector
$\mat{v}$. The notation $\lonenorm{\mat{v}}$ denotes the $\ell_1$-norm of the vector $\mat{v}$
and is defined as $\lonenorm{\mat{v}} \df \sum_i \left |(\mat{v})_i\right |$~. The
notation $\ltwonorm{\mat{v}}$ denotes the Euclidean norm of the vector $\mat{v}$ and is
defined as $\ltwonorm{\mat{v}} \df \sqrt{\sum_i \left |(\mat{v})_i\right |^2}$~.
 The notation $a\sim\N(\mu,\sigma^2)$ denotes a Gaussian random variable $a$ with mean $\mu$ and variance $\sigma^2$~.

 The notation
$\A\otimes\B=[(\A)_{ij}\B]$ (written in block matrix form, where
the pair $i,j$ spans the range of indexes of $\A$) denotes the
Kronecker product of $\A$ times $\B$

\subsection{Compressed Sensing}\label{sec:CS}

In the standard CS framework, introduced in \cite{candes2006nos}, a signal $\x\in\Ri^{N\times 1}$
 which has  a sparse representation in some basis $\Ps\in\Ri^{N\times N}$, \textit{i.e}:
\begin{equation*}
\x = \Ps \bm{\theta},\quad \lzeronorm{\bm{\theta}} = K,\quad K\ll N
\end{equation*}
can be recovered by a smaller vector $\y\in\Ri^{M\times 1}$, $K<M<N$, of linear measurements $\y = \Ph\x$, where $\Ph\in\Ri^{M\times N}$ is the \emph{sensing matrix}. The optimum solution, requiring at least $M = K+1$ measurements, would be
$$
\widehat{\bm{\theta}}=\arg\min_{\bm{\theta}}\lzeronorm{\bm{\theta}}\ \quad \text{s.t.}\quad \Ph\Ps\bm{\theta} = \y~.
$$
Since the $\ell_0$ norm minimization is a NP-hard problem,
one can resort to a linear programming reconstruction by minimizing
the $\ell_1$ norm
\begin{equation}\label{eq:CS_recovery}
\widehat{\bm{\theta}}=\arg\min_{\bm{\theta}}\lonenorm{\bm{\theta}}\ \quad \text{s.t.}\quad \Ph\Ps\bm{\theta} = \y~,
\end{equation}
provided that $M$ is large enough ($\sim K\log(N/K)$).

The same algorithm holds for signals which are not exactly sparse, but rather compressible, meaning that they (or their representation $\bm{\theta}$ in basis \Ps) can be expressed only by $K$ significant coefficients, while the remaining ones are (close to) zero.

When the measurements are noisy the $\ell_1$ minimization
with relaxed constraints is used for reconstruction:
\begin{equation}\label{eq:CS_recovery_relaxed}
\widehat{\bm{\theta}}=\arg\min_{\bm{\theta}}\lonenorm{\bm{\theta}}\ \quad \text{s.t.}\quad \ltwonorm{\Ph\Ps\bm{\theta} - \y} < \varepsilon~,
\end{equation}where $\varepsilon$ bounds the amount of noise in the
data.
It has been shown in \cite{baraniuk2008spr} that extracting the elements of \Ph\ at random from a Gaussian or Rademacher distribution (i.e., $\pm 1$ with the same probability), and, in general, from any Sub-Gaussian distribution, allows a correct reconstruction with overwhelming probability.

\section{Proposed Architectures}\label{sec:prop_alg}

\begin{figure*}[t]
 \centering
 \subfigure[]{\label{fig:bd_classic}\includegraphics[width=\columnwidth]{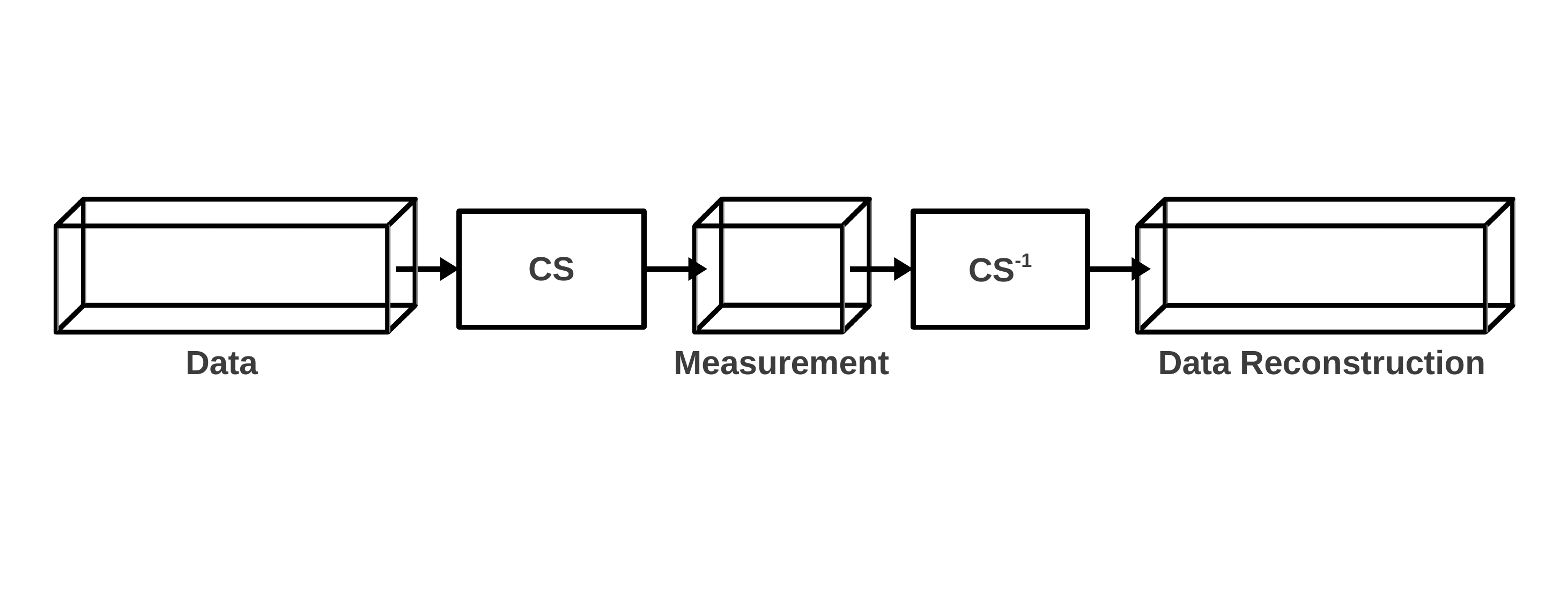}}
 \subfigure[]{\label{fig:bd_novel}\includegraphics[width=\columnwidth]{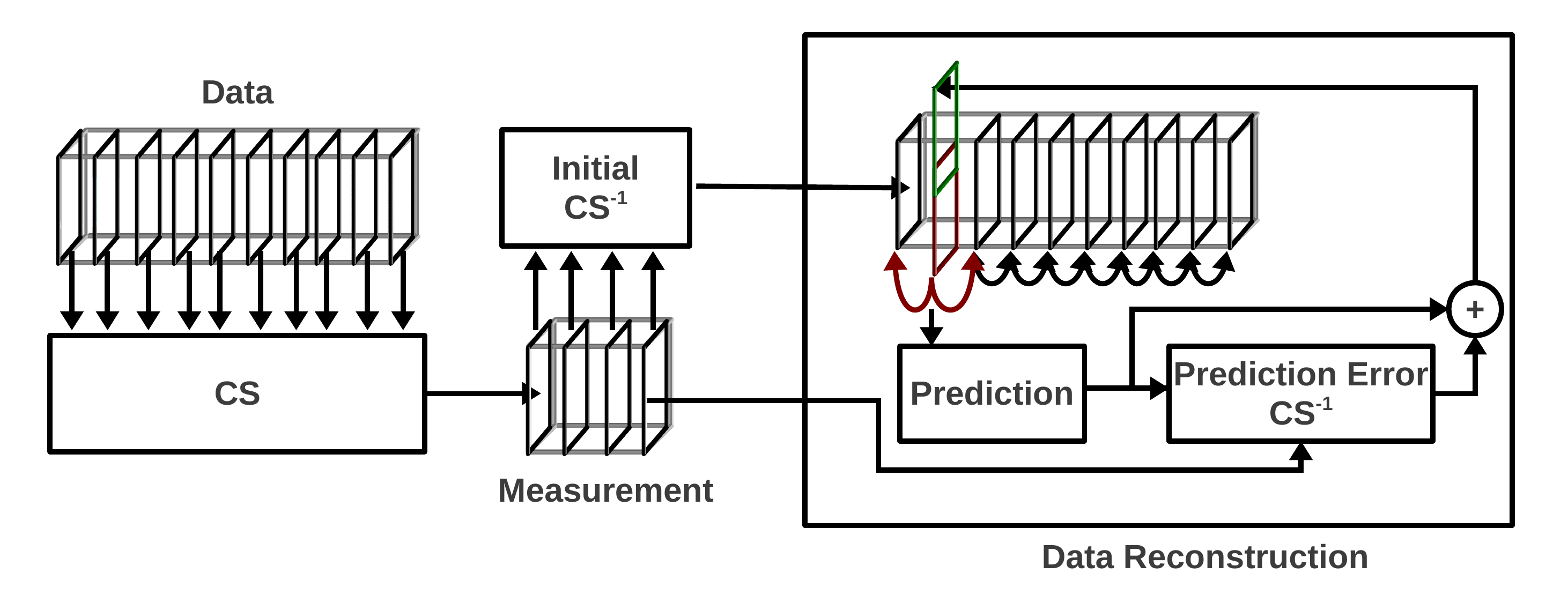}}
 \caption{Block Diagram of an architecture processing the whole signal (a) and of the novel iterative architecture (b).}
\label{fig:block_diagrams}
 \end{figure*}

In this paper, we propose a general framework for multidimensional signals, allowing to exploit the low complexity and universality of CS in the acquisition process, with a manageable complexity of the reconstruction algorithm.
Refer to Figure~\ref{fig:block_diagrams}, which depicts the canonical approach of considering the multidimensional signal as a single signal (\ref{fig:bd_classic}) and the architecture proposed in this paper (\ref{fig:bd_novel}), based on progressive scanning and iterative reconstruction.
Multidimensional signals are often captured in a progressive way, in a sequence of acquisitions corresponding to subsets of the coordinates. The principle is to acquire separately each signal dimension (or subsets of dimensions, considering them as single signals). Then, instead of reconstructing the whole set of measurements at once, as done with Kronecker Compressed Sensing \cite{duarte2009kronecker}, an iterative algorithm is applied to dimensions not involved in the measurement process. At each iteration, a linear prediction filter is used to aid the reconstruction process, measuring the prediction and reconstructing the prediction error only, which is supposed to be more compressible than the original signal. For example, if an hyperspectral image is acquired band by band, the iterative algorithm is applied on wavelength dimension. At each iteration, each band is predicted, acquired with the same sensing matrix used to acquire that band, the measurement of the predicted band is subtracted from the measurement of the band itself and the CS reconstruction is applied to this ``measurement prediction error'', only.  

In the following, we specialize the approach to 2D and 3D signals. We propose novel architectures for 2D and 3D acquisition and reconstruction, and the corresponding algorithms implementing the proposed architectures.

\subsection{2D signals}
According to typical progressive scanning approaches, like the ones used by commercial flatbed scanners or by remote sensing systems acquiring environmental pictures, an image is acquired by sensing $N_{\mathsf{COL}}$ pixels of each row in a progressive fashion, until $N_{\mathsf{ROW}}$ rows are acquired. Hence, the acquired image will result as a matrix of pixels of size $N_{\mathsf{ROW}} \times N_{\mathsf{COL}}$, which will be compressed (and, accordingly, decoded) using a conventional technique. This process requires the acquisition (and processing) of $N_{\mathsf{ROW}} N_{\mathsf{COL}}$ pixels. When $N_{\mathsf{ROW}}$ and  $N_{\mathsf{COL}}$ are large, processing of this huge amount of data may represent an issue, especially when dealing with low cost or low complexity devices.

For this reason, we propose a very simple acquisition scheme, based on CS linear measurements taken on each row, without any further processing. This reduces the amount of data to be acquired and processed. The reconstruction algorithm relies on linear prediction filters in order to  improve the quality of CS reconstruction, by correlating the measurements of adjacent rows in order to exploit their statistical dependencies during the reconstruction stage, largely improving over individual separate reconstruction.

\subsubsection{Acquisition}\label{sec:coding}

The image acquisition algorithm we propose, labelled as Algorithm~\ref{alg:2D_coding}, is very simple and consists in taking linear measurements of each row of the image in a progressive fashion. To minimize the risks of failures in the reconstruction side, a different sensing matrix \Ph\ is drawn for each row.

The image to be measured can be divided into $N_{\mathsf{ROW}}$ rows. For each row, $M$ linear measurements are taken, where $M < N_{\mathsf{COL}}$ and $N_{\mathsf{COL}}$ is the desired vertical resolution.

In summary, the scene we wish to acquire is represented by the matrix  $\X \in \Ri^{N_{\mathsf{ROW}} \times N_{\mathsf{COL}}}$. For each row of \X, we draw a matrix $\Ph^i \in \Ri^{M \times N_{\mathsf{COL}}}$ whose elements are Gaussian i.i.d. such that $(\Ph^i)_{kj}\sim\N(0,1/M)$, with $k = 1, \ldots, M$ and $j = 1, \ldots, N_{\mathsf{COL}}$. Then, we take $M$ linear measurements of $(\X)_i$ which will form the rows of the matrix of measurements $\Y \in \Ri^{N_{\mathsf{ROW}} \times M}$, namely
$$
\trasp{(\Y)_i} = \Ph^i\trasp{(\X)_i}
$$

\begin{algorithm}[t]
\caption{Acquisition algorithm for 2D signals}\label{alg:2D_coding}
\begin{algorithmic}[1]
\Require the image \X, $M$
\Ensure the measurement matrix \Y
\For{$i = 1$ to $N_{\mathsf{ROW}}$}
\State Draw $\Ph^i$ of size $M\times N_{\mathsf{COL}}$ s.t. $(\Ph^i)_{kj}\sim\N(0,1/M)$
\State$ \trasp{(\Y)_i} \gets \Ph^i\trasp{(\X)_i}$
\EndFor
\State \Return \Y
\end{algorithmic}
\end{algorithm}

A more complex algorithm, based on Compressed Sensing and able to capture spatial correlation in both directions (horizontal and vertical), could acquire in a single shot the whole image in a single measurement vector of length $M'$.
$$
\y' = \Ph'\vect{\X}~,
$$
where $\vect{\X} \in \Ri^{N_{\mathsf{ROW}}N_{\mathsf{COL}} \times 1}$, $\Ph' \in \Ri ^{M' \times N_{\mathsf{ROW}}N_{\mathsf{COL}}}$, $\y' \in \Ri^{M'\times 1}~.$

Even if this algorithm performed better than the one proposed here since the reconstruction would optimally exploit the correlation in 2 dimensions through a 2D transform matrix, it would require the solution of \eqref{eq:CS_recovery} for a vector of length $N=N_{\mathsf{ROW}}N_{\mathsf{COL}}$. For realistic values of $N_{\mathsf{ROW}}$ and $N_{\mathsf{COL}}$, the solution of \eqref{eq:CS_recovery} would be impossible to perform in reasonable time.
On the other hand, the proposed approach splits the problem into smaller (and hence tractable) subproblems. However, in doing so, it does not neglect the spatial correlation in vertical direction, which is modelled and employed in the reconstruction process through the use of linear prediction filters.

\subsubsection{Reconstruction}\label{sec:decoding}

A trivial reconstruction algorithm based on the acquisition scheme described in section \ref{sec:coding} would simply apply the $\ell_1$ reconstruction \eqref{eq:CS_recovery} to recover separately each line of \X\ given the corresponding $\Ph^i$ and $(\Y)_i$.

Instead, we propose an algorithm which iteratively improves the current estimate of \X\ by modelling statistical dependencies between adjacent lines. We label this Algorithm~\ref{alg:2D_decoding}. We count the iterations using the index $n$. The estimation of \X\ at iteration $n$ is denoted with $\X^{(n)}$.

In particular, the algorithm evaluates a first image reconstruction using some initialization strategy $\digamma(\Y, \Ph)$ (iteration $n=0$). Then, the iterations start. The intuition is as follows. For each row, if we are able to reliably predict it using the reconstruction of the upper and lower lines at previous iteration with some linear prediction filter $\mathsf{P(\cdot,\cdot)}$, obtaining $\x_\mathsf{P}$, we can compute the ``measurement'' $\y_\mathsf{P}$ of this prediction by applying matrix $\Ph^i$ to $\x_\mathsf{P}$. Then we calculate the prediction error in the linear measurement domain $\e_{\y}$ by subtracting this ``predicted measurement'' from the original measurement row $(\Y)_i$. The error $\e_{\y}$  will be then reconstructed using \eqref{eq:CS_recovery}, leading to a prediction error on the signal samples equal to $\e_{\x}$. Adding $\e_{\x}$ to $\x_\mathsf{P}$ provides a new estimate of \x. Since the new estimate is more accurate than the old one, the process can be repeated by estimating a new, more accurate prediction.  If the prediction of the row is accurate enough, the prediction error is going to be more compressible than the original vector. As a consequence, for an equal number of measurements, the $\ell_1$ reconstruction will yield lower MSE. 
We support this claim by numerical simulations, whose results are shown in Section~\ref{sec:pred_err_sparsity}.

Different initialization strategies are described in section ~\ref{sec:init_strat}, while prediction filters for 2D signals are described in section~\ref{sec:lp_2D}. Since \eqref{eq:CS_recovery} is a convex problem and the prediction filters we test are linear, the overall algorithm can be considered as a projection onto convex sets. This ensures the convergence of the algorithm to the intersection of the constraint sets (if any)  \cite{combettes1993foundations}.

\begin{algorithm}[t]
\caption{Reconstruction algorithm for 2D signals}\label{alg:2D_decoding}
\begin{algorithmic}[1]
\Require the measurement matrix \Y, the set of $\Ph^i$
\Ensure the estimation $\widehat{\X}$
\State $n \gets 0$
\State $\X^{(n)} \gets \digamma(\Y, \Ph)$
\Repeat
\State $n \gets n+1$
\For{$i = 1$ to $N_{\mathsf{ROW}}$}
\If{$i = 1$ or $i = N_{\mathsf{ROW}}$}
\State $\x_\mathsf{P} \gets \trasp{(\X^{(n-1)})_i}$
\Else
\State $\x_\mathsf{P} \gets \trasp{\mathsf{P}\left((\X^{(n-1)})_{i-1},(\X^{(n-1)})_{i+1}\right)}$
\EndIf
\State $\y_\mathsf{P} \gets \Ph^i\x_\mathsf{P}$
\State $\e_{\y} \gets \trasp{(\Y)_i} - \y_\mathsf{P}$
\State $\e_{\bm{\theta}} \gets  \arg\min_{\e}\lonenorm{\e}\ \quad \text{s.t.}\quad \Ph^i\Ps\e = \e_{\y} $
\State $\e_{\x} \gets \Ps\e_{\bm{\theta}}$
\State $\trasp{(\X^{(n)})_i} \gets \trasp{(\x_\mathsf{P} + \e_{\x})}$
\EndFor
\Until{Convergence is reached}
\State \Return $\X^{(n)}$
\end{algorithmic}
\end{algorithm}

\subsection{3D signals}

As has been said, acquisition of 3D hyperspectral images can be performed in different ways. Common to the various approaches are the signal dimensions, which determine the spatial and spectral resolution of the imaging system. The spectral resolution is given by the number of individual wavelengths $N_{\mathsf{BAND}}$ that the system is able to discriminate. In one possible approach, each wavelength is sensed individually, leading to different measurements for each spectral channel. In another approach, the system measures spatial-spectral slices individually, and different wavelengths are separated during the CS reconstruction process. In both cases, we assume that the user intends to acquire or reconstruct each spectral channel with a resolution of $N_{\mathsf{ROW}}\times N_{\mathsf{COL}}$ pixels. In both cases, we can represent the original data as a cube with two spatial and one spectral dimension, and interpret them either as a collection of $N_{\mathsf{BAND}}$ spectral channels, or as a collection of $N_{\mathsf{ROW}}$ spatial-spectral slices. In the following we take the first approach, and consider an hyperspectral image $\mathcal{F} \in \Ri^{N_{\mathsf{ROW}}\times N_{\mathsf{COL}}\times N_{\mathsf{BAND}}}$ as a collection of $N_{\mathsf{BAND}}$ spectral channels $\F^i = \mathcal{F}_{:,:,i}, i=1,\ldots,N_{\mathsf{BAND}}$, each consisting of a $N_{\mathsf{ROW}}\times N_{\mathsf{COL}}$ frame, i.e.
$$\mathcal{F}=[\F^1, \F^2, \ldots, \F^{N_\mathsf{BAND}}]~.$$

\subsubsection{Acquisition}

According to Algorithm~\ref{alg:3D_coding_band_unique}, for each spectral channel a collection
$\y_i\in\Ri^{M\times 1}$ of $M$ measurements is acquired as
$$
\y_i=\Ph^i\vect{\F^i}~,
$$
where each sensing matrix $\Ph^i\in\Ri^{M\times N_{\mathsf{ROW}} N_{\mathsf{COL}}}$ is taken as Gaussian i.i.d. and $M<N_{\mathsf{ROW}}N_{\mathsf{COL}}$. For simplicity, $M$ is taken as
the same value for all spectral channels. The measurements of all channels are then collected in the matrix \Y. This setting
is amenable to separate spatial reconstruction of each spectral
channel using a two-dimensional transform as sparsity domain.
However, we expect that separate spatial reconstruction does not
yield a sufficiently accurate estimate of the original image, since
it lacks modelling of spectral correlation, which is very strong for
hyperspectral images.

\begin{algorithm}[t]
\caption{Acquisition algorithm for 3D signals}\label{alg:3D_coding_band_unique}
\begin{algorithmic}[1]
\Require the hyperspectral image $\mathcal{F}$, $M$
\Ensure the measurement matrix \Y
\For{$i = 1$ to $N_{\mathsf{BAND}}$}
\State Draw $\Ph^i$ of size $M\times N_{\mathsf{ROW}}N_{\mathsf{COL}}$ s.t. $(\Ph^i)_{kj}\sim\N(0,1/M)$
\State $\F^i = \mathcal{F}_{:,:,i}$
\State$ \trasp{(\Y)_i} \gets \Ph^i\vect{\F^i}$
\EndFor
\State \Return \Y
\end{algorithmic}
\end{algorithm}

\subsubsection{Reconstruction}

The idea behind the iterative reconstruction is that, as in the 2D case, if we can
obtain a prediction of a spectral channel $\F^i$, e.g. applying the
operator  $\mathsf{P(\cdot,\cdot)}$ to channels $\F^{i-1}$ and
$\F^{i+1}$ of some initial reconstruction, then we can cancel out
the contribution of this predictor from the measurements of $\F^i$,
and reconstruct only the prediction error instead of the full
spectral channel. If the prediction filter is accurate, the prediction error
is expected to be more compressible than the full signal, and the
reconstruction will yield better results as shown in Section~\ref{sec:pred_err_sparsity}. In particular, the iterative procedure starts
from the initial reconstruction $\mathcal{F}^(0)$ of all spectral channels. At this stage, we do not specify how we generate such initial reconstruction,
which is generically denoted by $\digamma(\Y,\Ph)$ to indicate that it is computed from random projections $\Y$ and measurement matrices $\Ph$.
Then, for every channel we first obtain
$\F_\mathsf{P} = \mathsf{P}\left(\mathcal{F}^{(n-1)}_{:,:,i-1},\mathcal{F}^{(n-1)}_{:,:,i+1}\right)$.
After that, we compute prediction error measurements as $\e_{\y} =
\y_i- \Ph^i\vect{\F_\mathsf{P}}$, and we use $\e_{\y}$ to reconstruct the $i$-th
channel summing the CS reconstruction of $\e_{\y}$ to $\F_\mathsf{P}$. $\Ps_\mathsf{2D}$ is the 2D sparsity transform matrix. If $\bm{\theta} = \Ps_{\mathsf{ROW}}\X\trasp{\Ps_{\mathsf{COL}}}$ is the sparse representation of the 2D signal \X\ for some $\Ps_{\mathsf{ROW}}$ and $\Ps_{\mathsf{COL}}$, then $\vect{\X} = \Ps_\mathsf{2D}\vect{\bm{\theta}}$, where
\begin{equation}
\Ps_\mathsf{2D} = \left(\trasp{\Ps_{\mathsf{COL}}}\otimes\trasp{\Ps_{\mathsf{ROW}}}\right)~.
\label{eq:Psi2}
\end{equation} 
This process
is performed on all bands, and is iterated until convergence. Again, it is worth noting that since \eqref{eq:CS_recovery} is a convex problem and the prediction filter is linear, this
algorithm can be cast in terms of projections onto convex sets
\cite{combettes1993foundations}, guaranteeing convergence to the intersection of
the constraint sets (if not empty). The proposed iterative reconstruction scheme is shown in Algorithm \ref{alg:3D_decoding_band_unique}. 

\begin{algorithm}[tb]
\caption{Reconstruction algorithm for 3D signals}\label{alg:3D_decoding_band_unique}
\begin{algorithmic}[1]
\Require the measurement matrix \Y, the set of $\Ph^i$
\Ensure the estimation $\widehat{\mathcal{F}}$
\State $n \gets 0$
\State $\mathcal{F}^{(n)} = \digamma(\Y,\Ph)$
\Repeat
\State $n \gets n+1$
\For{$i = 1$ to $N_{\mathsf{BAND}}$}
\If{$i = 1$}
\State $\F_\mathsf{P} \gets \mathsf{P}\left(\cdot,\mathcal{F}^{(n-1)}_{:,:,i+1}\right)$
\ElsIf {$i = N_{\mathsf{BAND}}$}
\State  $\F_\mathsf{P} \gets \mathsf{P}\left(\mathcal{F}^{(n-1)}_{:,:,i-1},\cdot\right)$
\Else
\State  $\F_\mathsf{P} \gets \mathsf{P}\left(\mathcal{F}^{(n-1)}_{:,:,i-1},\mathcal{F}^{(n-1)}_{:,:,i+1}\right)$
\EndIf
\State $\y_\mathsf{P} \gets \Ph^i\vect{\F_\mathsf{P}}$
\State $\e_{\y} \gets \trasp{(\Y)_i} - \y_\mathsf{P}$
\State $\e_{\bm{\theta}} \gets  \arg\min_{\e}\lonenorm{\e}\ \quad \text{s.t.}\quad \Ph^i\Ps_\mathsf{2D}\e = \e_{\y} $
\State $\e_{\F} \gets \Ps_\mathsf{2D}\e_{\bm{\theta}}$
\State $\vect{\mathcal{F}^{(n)}_{:,:,i}} \gets (\vect{\F_\mathsf{P}} + \e_{\F})$
\EndFor
\Until{Convergence is reached}
\State \Return $\mathcal{F}^{(n)}$
\end{algorithmic}
\end{algorithm}

\begin{remark}
The proposed acquisition and reconstruction architecture for 3D signals can be applied in any of the 3 dimensions. For example, the acquisition can be performed for each spectral row (instead of spectral channels), and the reconstruction can be performed iterating over rows. The respective algorithms can be obtained by Algorithms~\ref{alg:3D_coding_band_unique} and \ref{alg:3D_decoding_band_unique} by properly rotating indexes and dimensions.
\end{remark}

\subsection{Linear Prediction Filters}

\subsubsection{Row prediction filters for 2D signals}\label{sec:lp_2D}

We consider here several linear prediction filters $\mathsf{P}\left((\X)_{i-1},(\X)_{i+1}\right)$ for 2D signals, looking for the one providing fastest convergence and best MSE performance. We denote as $\x_\mathsf{P}$ the result of the prediction.

Prediction filter labelled as \emph{P1} estimates the current\footnote{Here and in the following equations, we omit the index ($n$) denoting current iteration} line to be predicted as the average of the upper and lower lines:
$$
\x_\mathsf{P} = \frac{1}{2}\trasp{\left((\X)_{i-1} + (\X)_{i+1}\right)}
$$

Prediction filter labelled as \emph{P2} predicts each pixel of current line as the average of adjacent pixels of upper and lower lines
\begin{align}
(\x_\mathsf{P})_j & = \frac{1}{6}\left[(\X)_{i-1,j-1} + (\X)_{i-1,j} + (\X)_{i-1,j+1}\right. \nonumber\\
& + \left.(\X)_{i+1,j-1} + (\X)_{i+1,j} + (\X)_{i+1,j+1}\right]~.\nonumber
\end{align}

Finally, prediction filter labelled as \emph{P3} predicts each pixel of current line as the \emph{weighted} average of adjacent pixels of upper and lower lines. Weights depend on the distance from the pixel to be predicted, namely
\begin{align}
(\x_\mathsf{P})_j & = \left[a(\X)_{i-1,j-1} + b(\X)_{i-1,j} + a(\X)_{i-1,j+1}\right. \nonumber\\
& + \left.a(\X)_{i+1,j-1} + b(\X)_{i+1,j} + a(\X)_{i+1,j+1}\right]~,\nonumber
\end{align}
with $a = \frac{2-\sqrt{2}}{4}$ and $b = \frac{\sqrt{2}-1}{2}$.

In section~\ref{sec:num_res_2D}, we test the performance of the linear prediction filters $\mathsf{P}(\cdot,\cdot)$ and of the overall algorithm.

\subsubsection{Band prediction filter for 3D signals}\label{sec:lp_3D}

In the following we describe the linear prediction stage $\mathsf{P}(\cdot,\cdot)$ employed
 in Algorithm~\ref{alg:3D_decoding_band_unique}.
The prediction filter operates in a blockwise fashion. Prediction of spectral channel $i$
is performed dividing the channel into non-overlapping spatial blocks of size $16\times 16$
pixels. Each block is predicted from the spatially co-located block
in a reference spectral channel $l$ (typically the previous or next
band). Focusing on a single $16\times 16$ block, we denote by $f_{m,n,i}$
the pixel of an hyperspectral image in $m$-th line, $n$-th pixel,
and $i$-th band, with $m,n=0, \ldots, 15$, and $i=0, \ldots, N_{\mathsf{BAND}}-1$.

Samples $f_{m,n,i}$ belonging to the block are predicted from the
samples $\widehat{f}_{m,n,l}$ of the {\em reconstructed} reference
band. In particular, a least-squares estimator \cite{slyz2005block} is
computed over the block. First, a gain factor is calculated as
$\alpha=\frac{\alpha_N}{\alpha_D}$, with $\alpha_N=\sum\limits_{m,n}
[(\widehat{f}_{m,n,l}-\mu_{l}) (\widehat{f}_{m,n,i}-\mu_{i})]$ and
$\alpha_D=\sum\limits_{m,n}[(\widehat{f}_{m,n,l}-\mu_{l})^2]$.
$\mu_{i}$ and $\mu_{l}$ are the average values of the co-located
reconstructed blocks in bands $\F^i$ and $\F^l$. Then the predicted
values within the block are computed for all $m,n=0, \ldots, 15$ as
$\widetilde{f}^{(l)}_{m,n,i}=
\mu_{i}+\alpha(\widehat{f}_{m,n,l}-\mu_{l})$.

This one-step prediction filter is employed in such a way as to take full
advantage of the correlation between bands. In particular, the current band is
very correlated with its two adjacent bands, while the correlation
tends to decrease moving further away. Eventually, we define a
predictor for a block in the current band $\F^i$ as the average of two
predictors obtained from the previous and the next band.
$\widetilde{f}_{m,n,i}=(\widetilde{f}^{(i-1)}_{m,n,i}+\widetilde{f}^{(i+1)}_{m,n,i})/2$.
Hence, the prediction filter $\mathsf{P}\left(\cdot,\cdot\right)$ applies this predictor to the two adjacent reconstructed
spectral channels in a blockwise manner as
described above, yielding a predicted spectral channel $\F_\mathsf{P}$.
Exceptions are made for the first and last band, where only the
available previous/next band is used for the prediction.

\subsection{Initialization Strategies}\label{sec:init_strat}

\subsubsection{Separate reconstruction}

Separate reconstruction represents the trivial way for reconstructing signal acquired with progressive algorithms like Algorithm~\ref{alg:2D_coding} or Algorithm~\ref{alg:3D_coding_band_unique}. It simply consists in applying the $\ell_1$ reconstruction \eqref{eq:CS_recovery} to recover each separately acquired portion of the original signal (i.e. a line $(\X)_i$ of an image or the frame $\F^i$ of an hyperspectral image) independently from each other, given the corresponding $\Ph^i$ and $(\Y)_i$.

\subsubsection{Kronecker Compressed Sensing}\label{sec:kcs}

Given the above, we have investigated the possibility of implementing a more sophisticated reconstruction algorithm
 which allows the proposed scheme to achieve good performance even for low $M$, i.e., for high compression ratios.
 To this aim, we considered the simplified 3D reconstruction scheme proposed
in \cite{duarte2009kronecker, duarte2010kronecker}, where it is shown that Kronecker product matrices are a
natural way to generate sparsifying and measurement matrices for the application of
CS to multidimensional signals, resulting in a formulation that is denoted by Kronecker Compressive Sensing (KCS).
In KCS, starting from the assumption that the signal structure
along each dimension can be expressed via sparsity, Kronecker
product sparsity bases combine the structures for each signal dimension
into a single matrix and representation. This allows to obtain separable transforms matrices, thus maintaining
 the computational complexity to an acceptable level. Similarly, Kronecker
random product measurement matrices for multidimensional signals can be
implemented by performing a sequence of separate random measurements
obtained along each dimension.
Given the above, the application of  KCS to the problem at hand is straightforward. We describe the initialization based on KCS for the more general 3D signal case, which can be trivially specialized to the 2D case: the separate (band by band)
 random projections $ \trasp{(\Y)_i} = \Ph^i\vect{\F^i}$ can be used to get a reconstruction scheme which profitably exploits correlation
 in all dimensions by using a separable 3D Kronecker
product sparsity domain. More specifically, we consider DCT transforms for both spatial and spectral domains since
 DCT transform is better than other typical transforms used in CS (e.g. Wavelet
 transform) on small spatial crops, while a wavelet transform would arguably
provide better performance over a larger image. Accordingly, denoting by $\Ps_\mathsf{2D}$ and $\Ps_\lambda$ the DCT sparsifying
 operator for the spatial and spectral domain, respectively, reconstruction is given by
$$
\vect{\mathcal{F}^{(0)}} = \Ps_{\mathsf{3D}}\widehat{\bm{\theta}}~,
$$ 
where $\Ps_{\mathsf{3D}} = \Ps_\mathsf{2D} \otimes \Ps_\lambda$ and $\widehat{\bm{\theta}}$ can be obtained by means of linear program reconstruction
$$
\widehat{\bm{\theta}}=\arg\min_{\bm{\theta}}\lonenorm{\bm{\theta}}\ \quad \text{s.t.}\quad \Ph\Ps_{\mathsf{3D}}\bm{\theta} = \y~,
$$
 
where $\y=\vect{\trasp{\Y}}$ and \Ph\ is the \emph{block-diagonal} sensing matrix obtained as  $$\Ph=\diag{\Ph^1,\ldots,\Ph^{N_{\mathsf{BAND}}}}~.$$ The reconstructed set of images
 $\mathcal{F}^{(0)}$ can then be used as starting point for the iterative algorithm proposed in Algorithm \ref{alg:3D_decoding_band_unique}.

\subsection{Complexity}

We explain here the complexity reduction obtained using Algorithms~\ref{alg:2D_decoding} and \ref{alg:3D_decoding_band_unique} instead of the standard CS reconstruction algorithm, processing the 2D/3D signal as a whole. We specialize the discussion to 2D and 3D signals, but it can be easily extended to any multidimensional signal. For an $N_{\mathsf{ROW}}\times N_{\mathsf{COL}}$ image, the standard CS reconstruction algorithm has an $O(N_{\mathsf{ROW}}^3 N_{\mathsf{COL}}^3)$ complexity. Our algorithm performing $N_{\mathsf{ITER}}$ iterations has an $O(N_{\mathsf{ITER}} N_{\mathsf{ROW}} N_{\mathsf{COL}}^3)$ complexity, with, usually, $N_{\mathsf{ITER}}\ll N_{\mathsf{ROW}}, N_{\mathsf{COL}}$. Hence, the complexity gain that can be obtained is $\sim O(N_{\mathsf{ROW}}^2)$. For 3D signals the same considerations hold. In this case, the gain will be $\sim O(N_{\mathsf{BAND}}^2)$.

\section{Numerical Results}\label{sec:num_res}
\subsection{2D images}\label{sec:num_res_2D}
\subsubsection{Choice of the Prediction Filter}

\begin{figure}
\centering
\includegraphics[width=1.1\columnwidth]{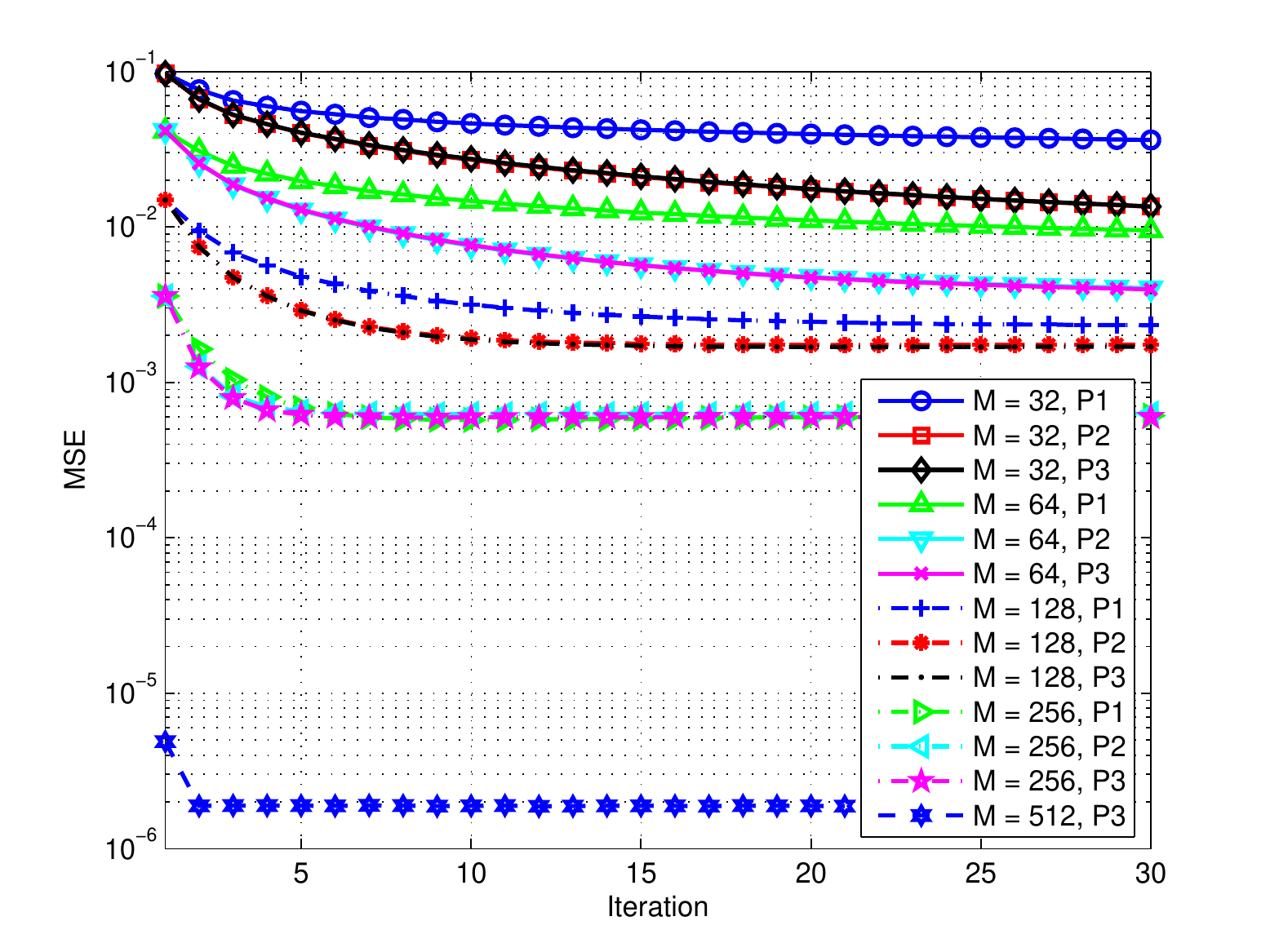}
\caption{{Test of different prediction filters on \emph{lena} image.}}
\label{fig:lena}
\end{figure}

First, we start by seeking the linear prediction filter $\mathsf{P}\left((\X)_{i-1},(\X)_{i+1}\right)$ providing fastest convergence and best MSE performance. For this test, we use the standard \emph{lena} black and white image of size $512 \times 512$. $M$ takes the values of $32, 64, 128, 256$ and the transform matrix $\Ps$ is the DCT matrix. We denote as $\x_\mathsf{P}$ the result of the prediction.

Fig.~\ref{fig:lena} shows the MSE performance of the overall system for different values of $M$ and using the prediction filters described above. Results show that the convergence is reached for each value of $M$. The bigger $M$, the faster the convergence and the smaller is the MSE at convergence. In any case, it can be noticed that the best performance is obtained for each value of $M$ using prediction filter labelled as \emph{P3}, i.e. the weighted average. Hence, we will use this prediction filter in our further tests, omitting to mention it from now on.

For $M=64$, the MSE obtained with separately recovered lines is $4.16\cdot 10^{-2}$. After 30 iterations, an MSE of $3.96\cdot 10^{-3}$ is obtained, with a gain of 10.2 dB. The convergence in this case is quite slow, but the MSE is decreased as much as one order of magnitude. Faster convergence is obtained with $M=128$, as after 15  MSE is decreased from $1.49\cdot 10^{-2}$ to $1.72\cdot 10^{-3}$, with a gain of 9.38 dB. Finally, with $M=256$ the MSE decreases from $3.59\cdot 10^{-3}$ to $6.18\cdot 10^{-4}$ in 5 iterations only, with a gain of 7.64 dB.

Finally, Fig.~\ref{fig:lena} shows as a reference the performance of a ``non compressive'' system. It can be noticed that the convergence is reached in only 1 iteration, leaving a residual MSE of $2\cdot 10^{-6}$ due to unpredictable signal components.

\subsubsection{Prediction error compressibility}\label{sec:pred_err_sparsity}
\begin{figure}
\centering
\includegraphics[width=1.1\columnwidth]{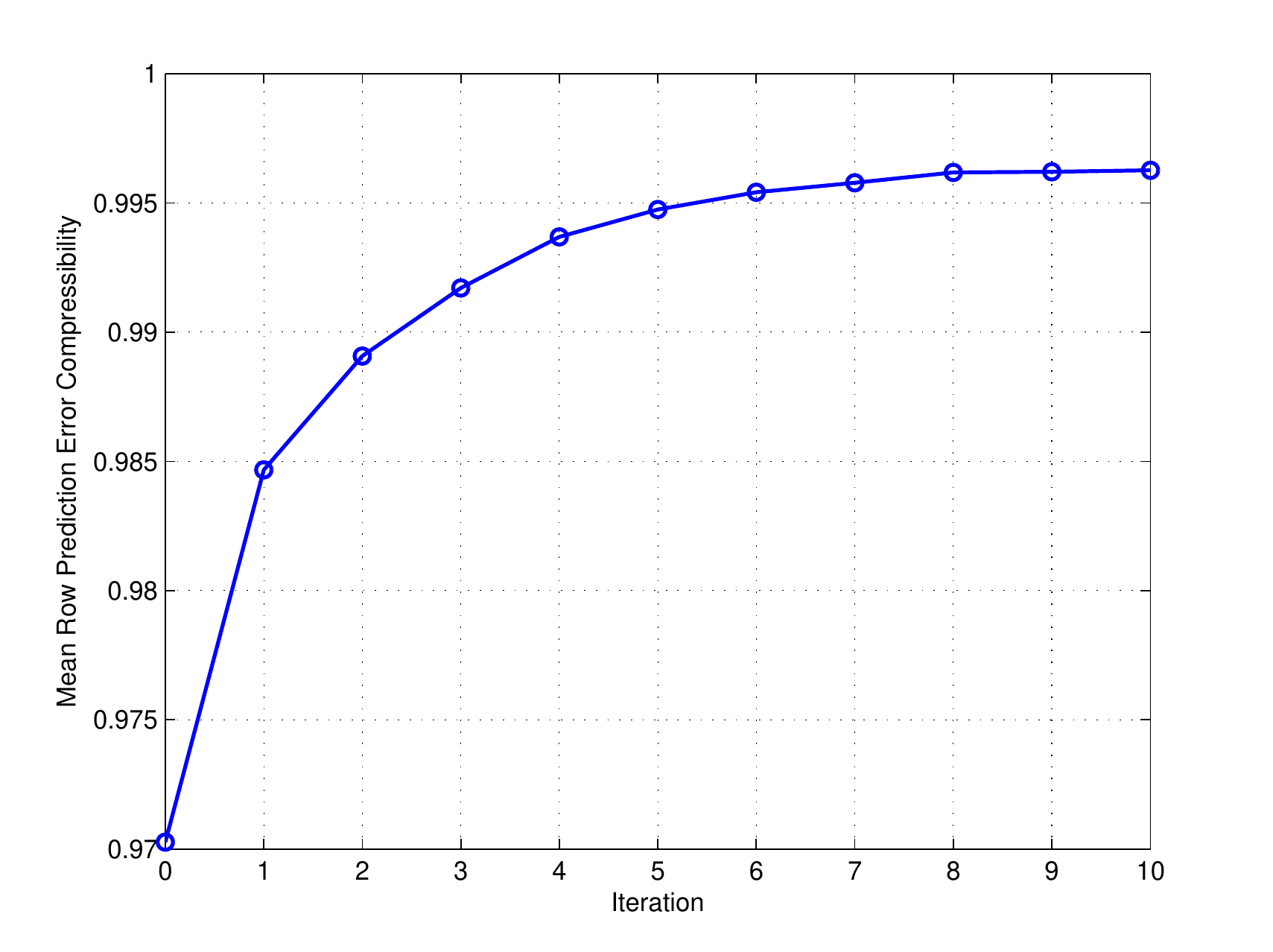}
\caption{Mean row compressibility of \emph{lena} image. $M=128$.}
\label{fig:lena_sparsity}
\end{figure}
In this section, we show results supporting the claim that accurate prediction leads to prediction errors which are more compressible than the original signal. Fig.~\ref{fig:lena_sparsity} shows the mean row compressibility of prediction error of the same image as previous section, measured at each of the first 10 iterations of the reconstruction algorithm. The data at 0-th iteration corresponds to the compressibility of the original image. With the term \emph{compressibility} here we mean the fraction of coefficients of the DCT of a row of the prediction error (or of the original image) below a certain threshold $\theta_\mathsf{TH}$, averaged over the rows. The threshold, evaluated for each row and at each iteration in order to take into account norm fluctuations, is computed as $\theta_\mathsf{TH} = 5\frac{\lonenorm{\bm{\theta}_i}}{N_\mathsf{COL}}$, where $\bm{\theta}_i$ is the DCT of the current row.

Fig.~\ref{fig:lena_sparsity} supports the claim that the prediction error is more compressible than the original signal and that it gets more and more compressible along iterations.

\subsubsection{Flatbed Scanner}

\begin{figure}[t]
 \centering
 \subfigure[Constellation]
   {\label{fig:constellation}\includegraphics[width=0.45\columnwidth]{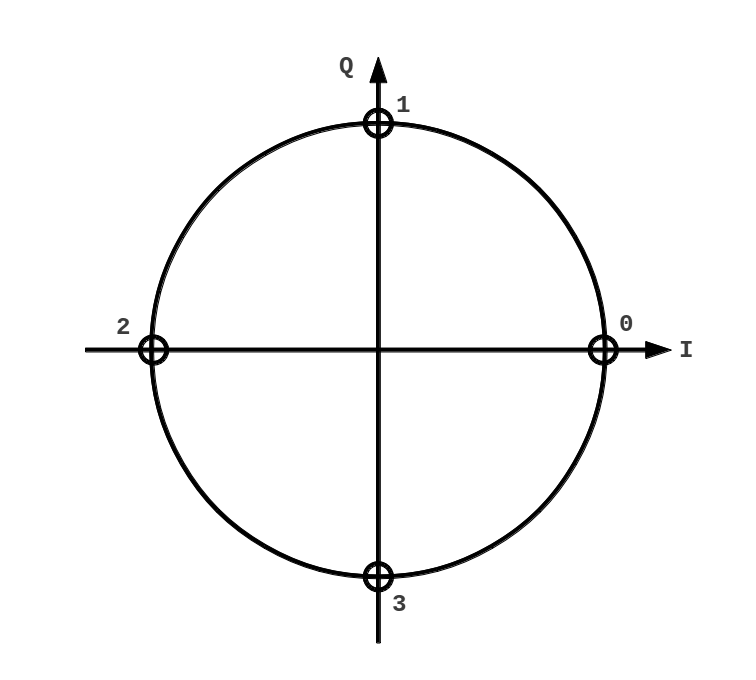}}
 \subfigure[Trellis]
   {\label{fig:trellis}\includegraphics[width=0.45\columnwidth]{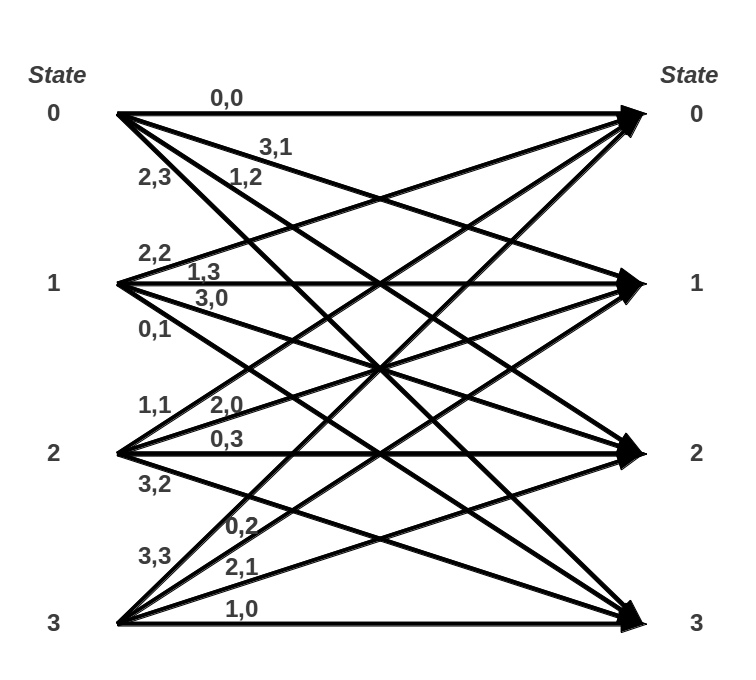}}
\subfigure[Block Diagram]
   {\label{fig:block_diagram}\includegraphics[width=0.9\columnwidth]{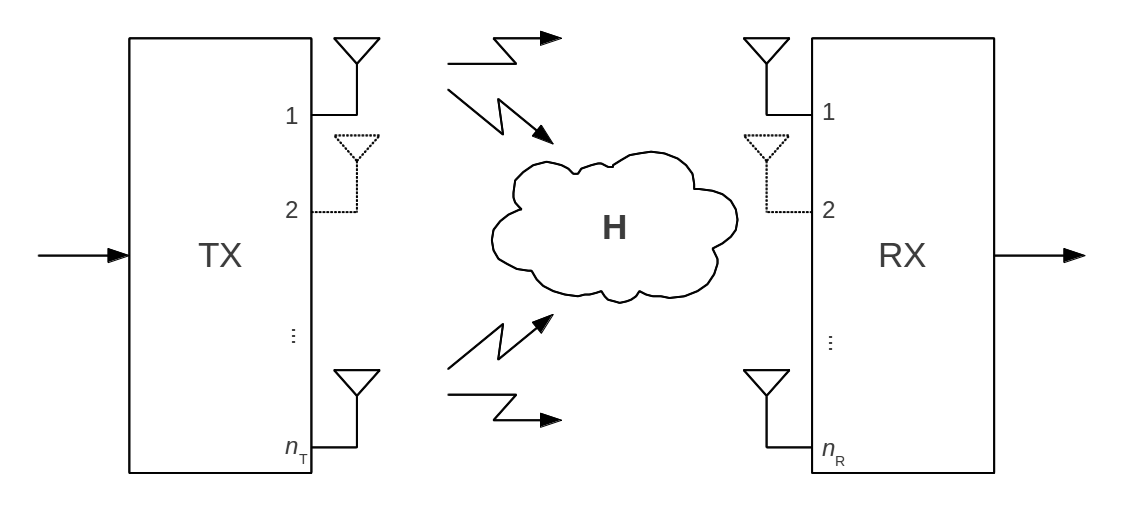}}
   \subfigure[Sample Text]
   {\label{fig:loremIpsum}\includegraphics[width=0.6\columnwidth]{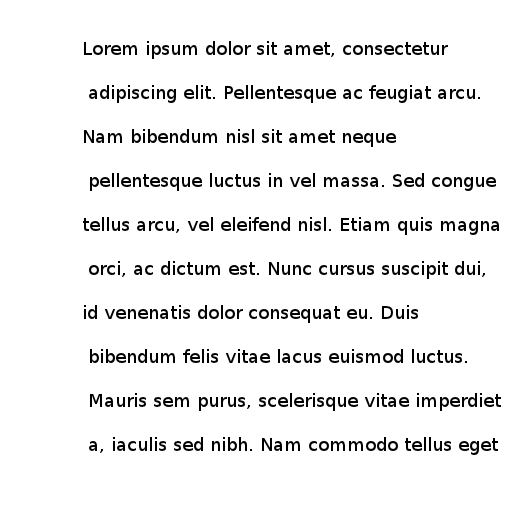}}

 \caption{{The graphics used as test image for \emph{flatbed scanner} scenario.}}
 \label{fig:graphs_bmp}
 \end{figure}

In this section, we apply our algorithm to images suitable to a \emph{flatbed scanner} scenario. These are black and white graphics and text, and are depicted in Fig.~\ref{fig:graphs_bmp}. $M$ takes the values of 8, 16, 32, 64, 128, 256 and (where possible) 512. Since they all have a completely white background (representing paper), they can be considered sparse in the pixel domain. Hence, the matrix \Ps\ is the identity matrix of size $N_{\mathsf{COL}}$, namely $\mat{I}_{N_{\mathsf{COL}}}$.

Fig.~\ref{fig:constellation} is the simplest graphic, representing a QPSK constellation. Fig.~\ref{fig:trellis} represents a slightly more complicated (hence, less sparse) graphic, the trellis of a convolutional code. Fig.~\ref{fig:block_diagram} is a larger figure representing a generic block diagram. Finally, fig.~\ref{fig:loremIpsum} depicts a sample of generic text.

Table~\ref{tab:scanner_res} reports the results obtained using the proposed algorithm. The table shows, for each image, the initial MSE (obtained using separate CS reconstruction of each line), the MSE the algorithm converges to, the performance gain, and the number of iterations necessary to reach convergence. Figures confirm the results obtained in the previous section. The more measurements are taken, the faster is the convergence and the lower is the MSE that can be obtained when the algorithm has converged. When the picture is very sparse, it is possible to obtain a reduction of one order of magnitude, while when the picture is less sparse the contribution of Compressed Sensing is weaker, but still a reduction of about 50\% in MSE can be obtained.

\begin{table}[t]
\caption{{MSE and convergence results on sample graphics}}
\centering
\begin{tabular}{|c|c|c|c|c|}
\hline
$M$ & init. MSE & conv. MSE & gain (dB) & steps \\
\hline
\multicolumn{5}{|c|}{\emph{Constellation} ($680\times576$)} \\
\hline
64 & $2.08\cdot 10^{-2}$ & $7.86\cdot 10^{-3}$ & 4.23 & 18 \\
\hline
128 & $9.99\cdot 10^{-3}$& $3.05\cdot 10^{-3}$ & 5.15 & 14\\
\hline
256 & $4.72\cdot 10^{-3}$ & $7.25\cdot 10^{-4}$ & 8.14 & 10 \\
\hline
\multicolumn{5}{|c|}{\emph{Trellis} ($680\times576$)} \\
\hline
64 & $8.56\cdot 10^{-2}$ & $3.98\cdot 10^{-2}$ & 3.33 & 18 \\
\hline
128 & $7.45\cdot 10^{-2}$& $2.00\cdot 10^{-2}$ & 5.71 & 13\\
\hline
256 & $3.39\cdot 10^{-2}$ & $6.51\cdot 10^{-3}$ & 7.17 & 8 \\
\hline
\multicolumn{5}{|c|}{\emph{Block Diagram} ($529\times1123$)} \\
\hline
64 & $8.38\cdot 10^{-3}$ & $7.02\cdot 10^{-3}$ & 0.77 & 7 \\
\hline
128 & $5.79\cdot 10^{-3}$& $3.97\cdot 10^{-3}$ & 1.64 & 6\\
\hline
256 & $2.79\cdot 10^{-3}$ & $1.66\cdot 10^{-3}$ & 2.25 & 5 \\
\hline
512 & $1.23\cdot 10^{-3}$ & $4.71\cdot 10^{-4}$ & 4.17 & 5 \\
\hline
\multicolumn{5}{|c|}{\emph{Sample Text} ($512\times512$)} \\
\hline
64 & $6.40\cdot 10^{-2}$ & $4.59\cdot 10^{-2}$ & 1.44 & 10 \\
\hline
128 & $5.46\cdot 10^{-2}$& $2.99\cdot 10^{-2}$ & 2.62 & 7\\
\hline
256 & $3.39\cdot 10^{-2}$ & $1.41\cdot 10^{-2}$ & 3.81 & 4 \\
\hline
\end{tabular}
\label{tab:scanner_res}
\end{table}

\subsubsection{Remote Sensing}

\begin{figure}
 \centering
 \includegraphics[scale=1]{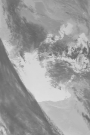}
 \caption{{The AIRS sensor gran 9 hyperspectral image, $600$-th band.}}
 \label{fig:airs_bmp}
 \end{figure}

To test the performance of the proposed scheme in a \emph{remote sensing} scenario, we use a spectral band extracted from hyperspectral image ``granule 9'' of the AIRS sensor. AIRS is an ultraspectral sounder with 2378 spectral channels, used to create 3D maps of air and surface temperature. The spatial size is $N_{\mathsf{COL}} = 90$ pixels and $N_{\mathsf{ROW}} = 135$ lines. The dataset consists in the raw output of the detector, without any processing, calibration or denoising applied. We choose the $600$-th band, which is depicted in Fig.~\ref{fig:airs_bmp}, but very similar results have been obtained with other bands and are omitted for brevity. $M$ takes the values of 8, 16, 32 e 64. The sparsity basis \Ps\ is the DCT.

Table~\ref{tab:airs9_600_results} (Basic Algorithm) summarizes the results obtained applying the proposed algorithm to the $600$-th band of the test image. Results show that with $M=8$ and $M=16$ the convergence is very slow and is not reached after 20 iterations. On the other hand, when $M=32$ the convergence is obtained after 10 iterations (reducing from $2.17\cdot 10^{-2}$ to $1.56\cdot 10^{-3}$, with a gain of 11.4 dB), while taking $M=64$ measurements per row implies the convergence after 4 steps only (with MSE reduction from $2.89\cdot 10^{-3}$ to $3.40\cdot 10^{-4}$ and a gain of 9.29 dB).
\begin{table}[t]
\caption{{MSE and convergence results on AIRS sensor image}}
\centering
\begin{tabular}{|c|c|c|c|c|}
\hline
$M$ & init. MSE & conv. MSE & gain (dB) & steps \\
\hline
\multicolumn{5}{|c|}{Basic algorithm} \\
\hline
8 & $2.40\cdot 10^{-1}$ & $2.63\cdot 10^{-2} $ (20-th it.) & 9.6 & 20+ \\
\hline
16 & $9.57\cdot 10^{-2}$ & $5.21\cdot 10^{-3}$ (20-th it.) & 12.6 & 20+ \\
\hline
32 & $2.17\cdot 10^{-2}$& $1.56\cdot 10^{-3}$ & 11.4 & 10\\
\hline
64 & $2.89\cdot 10^{-3}$ & $3.40\cdot 10^{-4}$ & 9.29 & 4 \\
\hline
\multicolumn{5}{|c|}{Kronecker improved algorithm} \\
\hline
8 & $4.60\cdot 10^{-3}$ & $3.80\cdot 10^{-3}$ & 0.83 & 7 \\
\hline
16 & $2.62\cdot 10^{-3}$ & $2.02\cdot 10^{-3}$ & 1.13 & 5 \\
\hline
32 & $1.22\cdot 10^{-3}$& $9.73\cdot 10^{-4}$ & 0.98 & 3\\
\hline
64 & $2.95\cdot 10^{-4}$ & $2.64\cdot 10^{-4}$ & 0.48 & 1 \\
\hline
\end{tabular}
\label{tab:airs9_600_results}
\end{table} 

As a term of comparison, we report here the MSE performance of a simple reconstruction algorithm named \emph{Orthogonal Matching Pursuit} \cite{tropp2007signal}, whose complexity is linear in the number of samples of the original signal ($N_\mathsf{ROW}N_\mathsf{COL}$ in this case). We acquire and reconstruct the entire image as a whole using $M=32\cdot N_\mathsf{ROW}$ and $M=64\cdot N_\mathsf{ROW}$ measurement, to be compared with the performance of our algorithm with $M=32$ and $M=64$, respectively. For $M=32\cdot N_\mathsf{ROW}$, we obtain an MSE of $1.9\cdot 10^{-3}$, while for $M=64\cdot N_\mathsf{ROW}$ we obtain an MSE of $1.7\cdot 10^{-3}$. Hence, our algorithm with $M=32$ performs 3 dB better than OMP with the same total amount of measurements, while with $M=64$ the gain is 8 dB.

\subsubsection{Improving performance with Kronecker Compressed Sensing}

An improvement to the performance of the algorithm is obtained using \emph{Kronecker Compressed Sensing} (KCS), described in section~\ref{sec:kcs}, into the algorithm we showcase. Hence, we use KCS to initialize the iterative algorithm we propose in this paper (instead of separate linewise reconstruction) and apply it to the \emph{remote sensing} scenario. The performance of this modified version of the algorithm are reported in Table.~\ref{tab:airs9_600_results} (Kronecker improved algorithm). The figures show two effects. First, the initial MSE is much lower than in the separate reconstruction case. This gain can be noticed in particular when $M$ is small and is due to the better performance of KCS reconstruction with respect to separate reconstruction; second, the iterative algorithm slightly improves the overall performance and converges in very few steps. This is due to the fact that KCS captures also correlation in vertical direction, making the contribution of each iteration less effective.

\begin{figure}
\centering
\includegraphics[width=1.1\columnwidth]{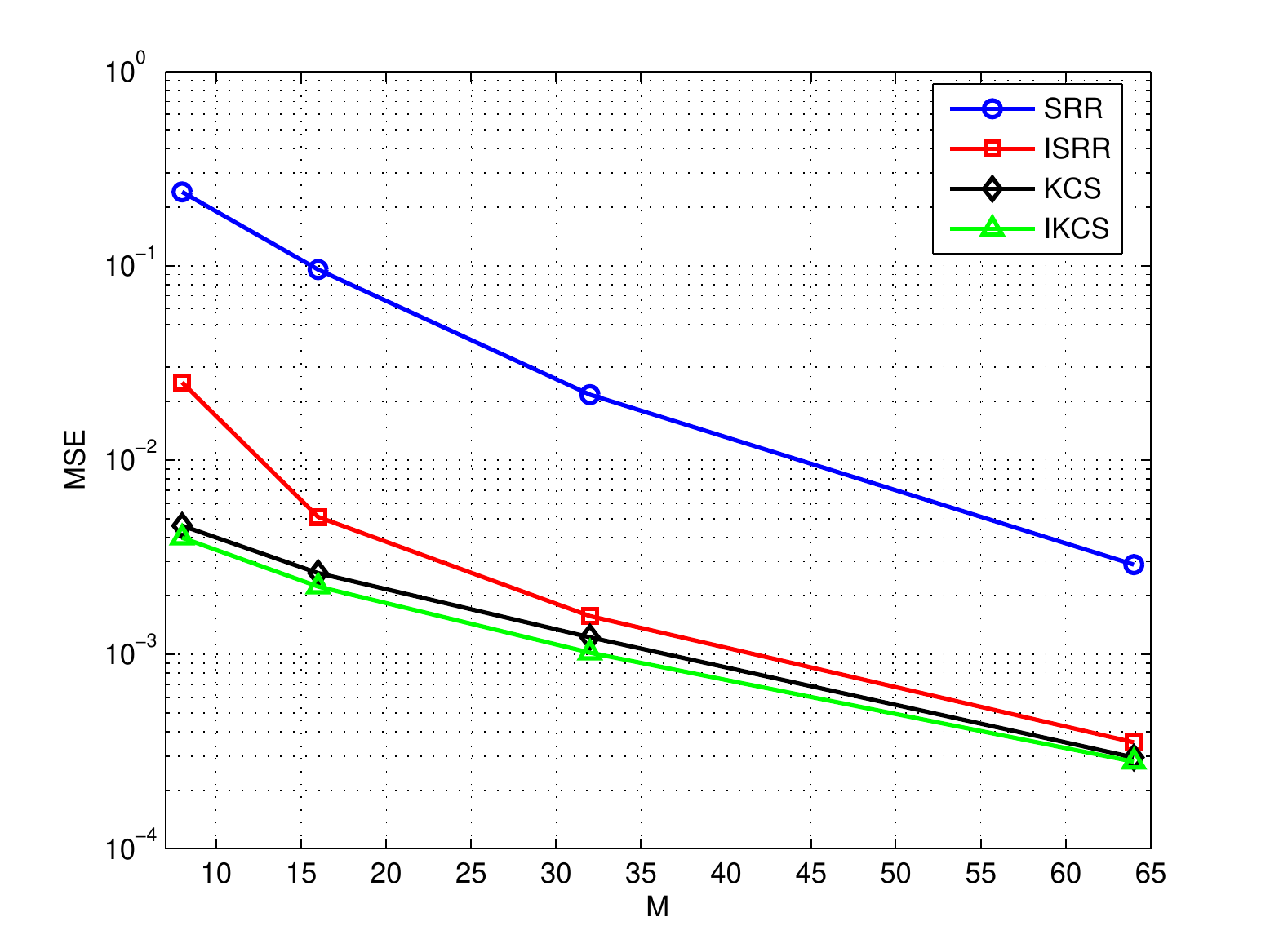}
\caption{{Performance comparison of proposed algorithms vs. $M$.}}
\label{fig:overall}
\end{figure}

Figure~\ref{fig:overall} summarizes the best MSE performance obtained by Separate Row Reconstruction (SRR), our Iterative algorithm initialized with Separate Row Reconstruction (ISRR), the Kronecker Compressed Sensing (KCS) and our Iterative algorithm with KCS initialization (IKCS) vs. the number of measurements $M$. Best performing algorithms are the ones implementing KCS. Plain KCS shows a gain of 7.37 dB over ISSR when $M=8$, and 1.10 dB when $M=32$. When using IKCS, roughly 1 dB of additional gain can be obtained with very few iterations.

\subsection{3D signals}\label{sec:num_res_3D}

We report reconstruction results on a few scenes that are used as
reference for onboard lossy compression in the ``multispectral and
hyperspectral data compression" working group of the Consultative
Committee for Space Data Systems (CCSDS), namely scene {\em sc0} of
AVIRIS (Yellowstone) and granule 9 ({\em gran9}) of AIRS. AVIRIS is
a spectrometer with 224 bands, and the size of this image is 512
lines and 680 pixels. AIRS has already been described in previous sections.
Because of the complexity of the reconstruction process and the
large amount of data, we do not use the complete images, but rather
a 32x32 spatial crop with all spectral channels.
Both are {\em raw} images, i.e. they are the output of the detector,
with no processing, calibration or denoising applied. These images
are noisier than the corresponding processed images, but more
realistic for application to onboard sensors.

\subsubsection{Preliminary experimental analysis with initial separate 2D reconstruction}

\begin{figure}
\begin{center}
 \includegraphics[width=0.5\textwidth]{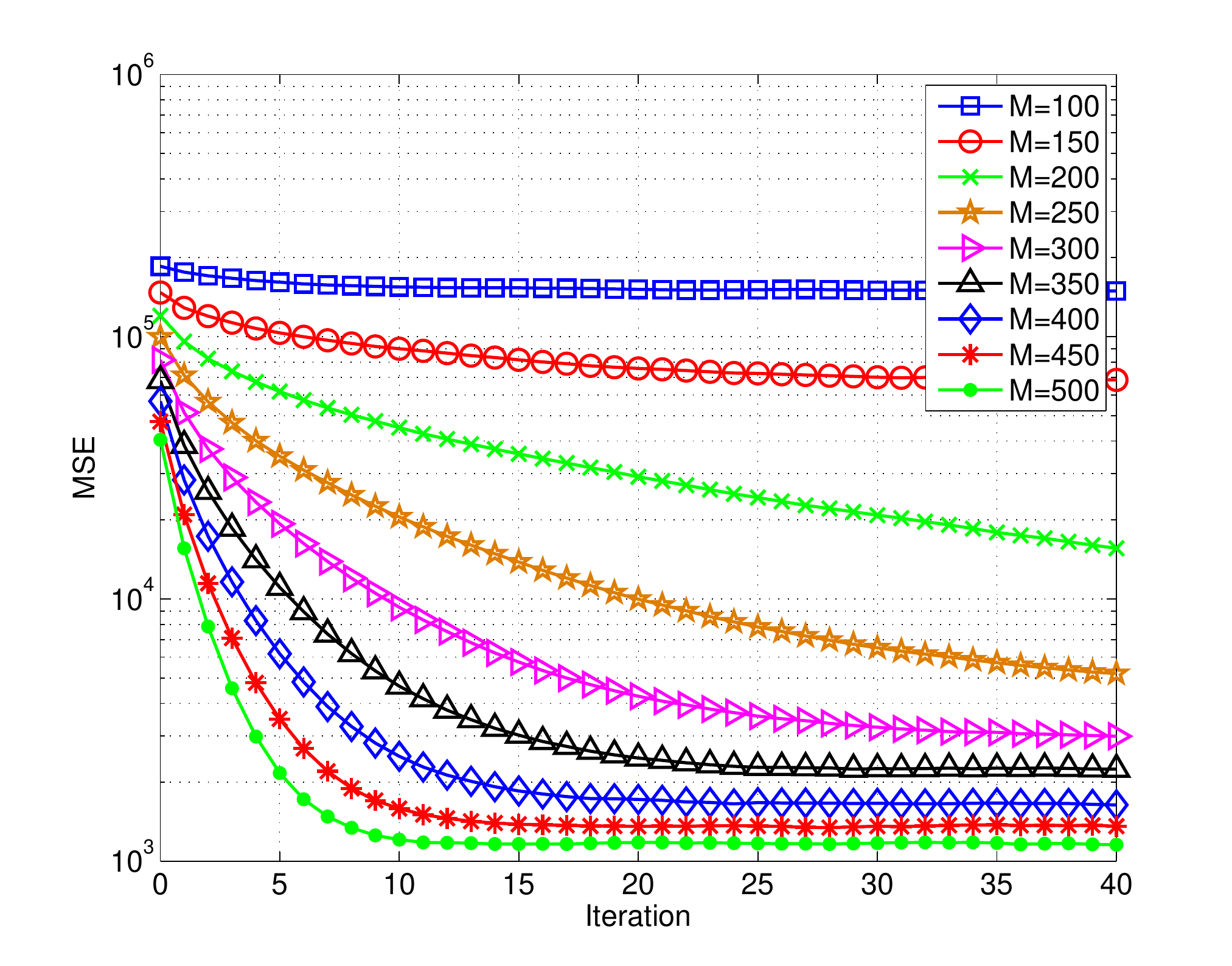} \\
\caption{AVIRIS: Reconstruction MSE of Algorithm \ref{alg:3D_decoding_band_unique} as a function of the number of iterations with initial separate 2D reconstruction.}
\label{fig:aviris_2D_iterations}
\end{center}
\end{figure}

We have carried out some experiments to preliminarily assess the validity of the proposed algorithm when the initial reconstruction images are computed using separate 2D DCT transforms band by band.
 In particular, Figure \ref{fig:aviris_2D_iterations} shows the MSE behaviour experienced on AVIRIS
 images as a function of the number of iterations for different values of the number of projections $M$.
 A similar behaviour is observed for AIRS images. Note that for medium to high $M$,
 iterations are effective in reducing MSE, e.g., for $M > 400$ the proposed
algorithm improves the MSE up to a factor of 35 with respect to the initial reconstruction.
 Moreover, convergence to the minimum attainable MSE  is obtained in a relatively small number of iterations.
For lower $M$, convergence is slower and MSE reduction is less effective.
 In particular, for very low $M$, e.g. for $M = 100$
 convergence
 is very slow and MSE reduction is negligible. In essence, the algorithm shows a threshold behaviour with respect to the
 initial reconstructed images: a poor initial reconstruction prevents the iterative algorithm
 to improve the MSE while if the initial reconstruction's MSE falls below a minimum threshold, the improvement
 is remarkable and convergence very fast.

\subsubsection{Improving initial reconstruction by means of Kronecker CS}
\label{sec:3Dalg}

\begin{figure}
\begin{center}
 \includegraphics[width=0.5\textwidth]{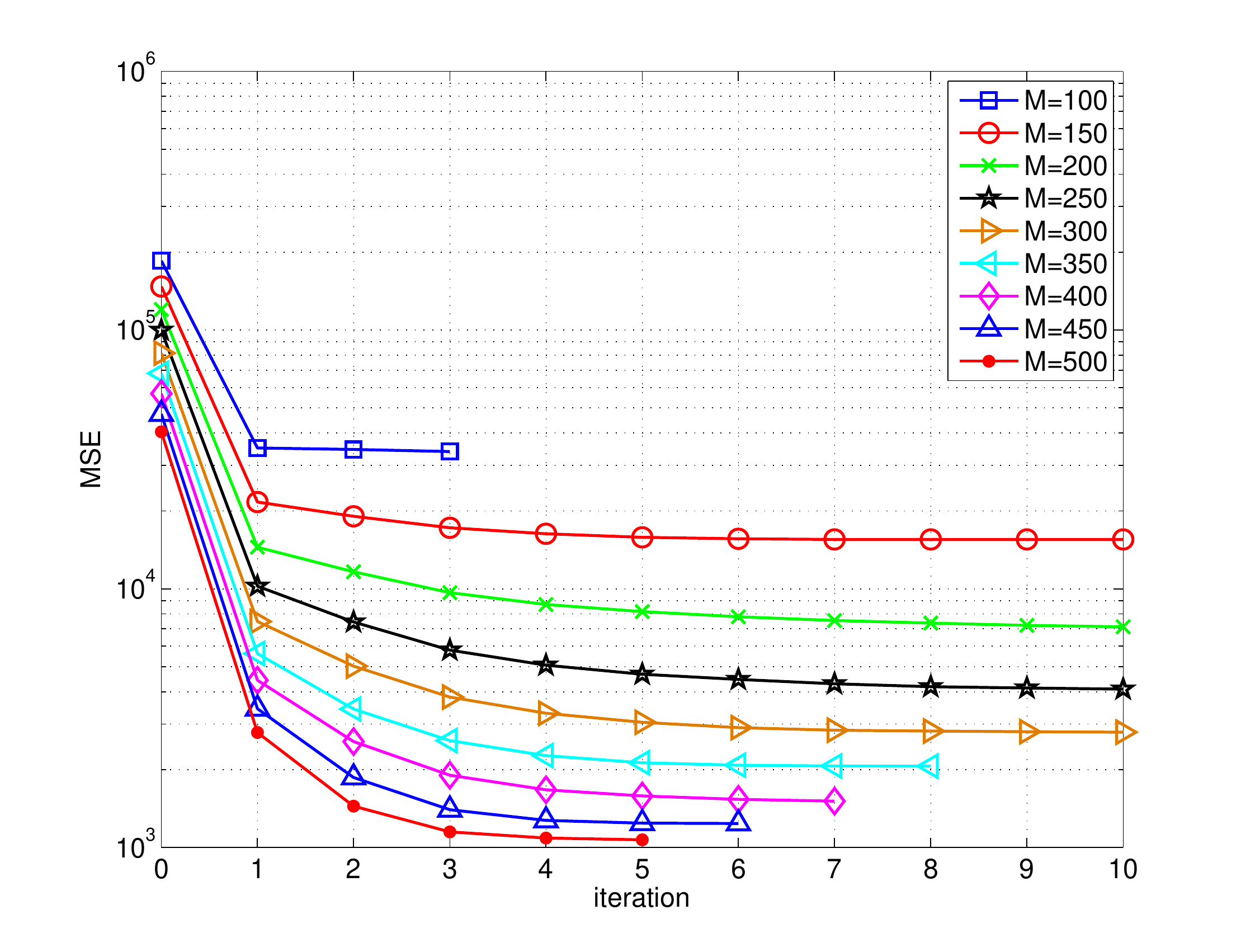} \\
\caption{AVIRIS: Reconstruction MSE of Algorithm \ref{alg:3D_decoding_band_unique} with 3D Kronecker starting point, as a function of the number of iterations.}
\label{fig:aviris3DIterations}
\end{center}
\end{figure}

 To assess the effectiveness of such an approach, in Figure \ref{fig:aviris3DIterations} we show the MSE behaviour experienced on AVIRIS
 images as a function of the number of iterations, for different $M$, when the starting point of the iterative scheme proposed in Algorithm \ref{alg:3D_decoding_band_unique} is obtained through Kronecker 3D reconstruction. Comparing with Figure \ref{fig:aviris_2D_iterations} it can be observed that, as expected, the MSE starting point is much lower and convergence is achieved in few iterations. Moreover, despite 3D Kronecker reconstruction already exploits correlation in the spectral domain, the proposed iterative algorithm still allows to improve the MSE up to a factor of 3 with respect to the initial reconstruction.
 In the next section, we describe in more details the experiments we conducted to evaluate the performance of the two proposed reconstruction schemes,
 namely iterative compressed sampling (ICS) and
 Kronecker-iterative compressed sampling (KICS), which are both based on the iterative procedure described
 in Algorithm \ref{alg:3D_decoding_band_unique}, with the initial point computed by means of separate reconstruction and KCS, respectively.

\subsubsection{Spectral channel analysis}

We compare results of the proposed ICS and KICS with
those obtained through separate spatial reconstruction (S2D) of each spectral
channel and through 3D KCS.
 The reconstruction algorithm for the iterative schemes is run for 40
iterations, with several values of $M$.

\begin{figure}
\begin{center}
 \includegraphics[width=0.49\textwidth]{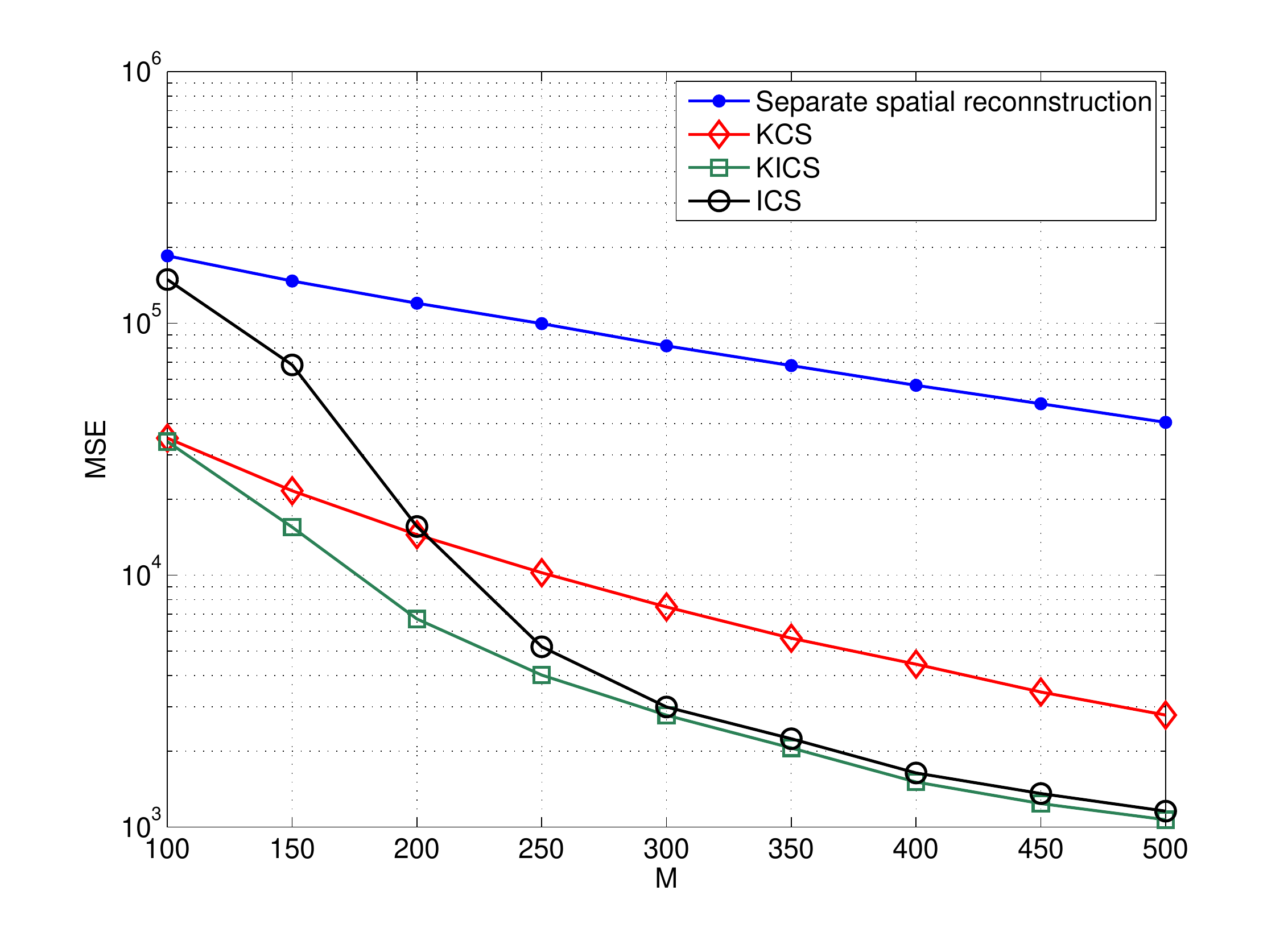} \\
\caption{AVIRIS: Reconstruction MSE vs. $M$ for different schemes.}
\label{fig:comparison_aviris}
\end{center}
\end{figure}

\begin{figure}
\begin{center}
 \includegraphics[width=0.49\textwidth]{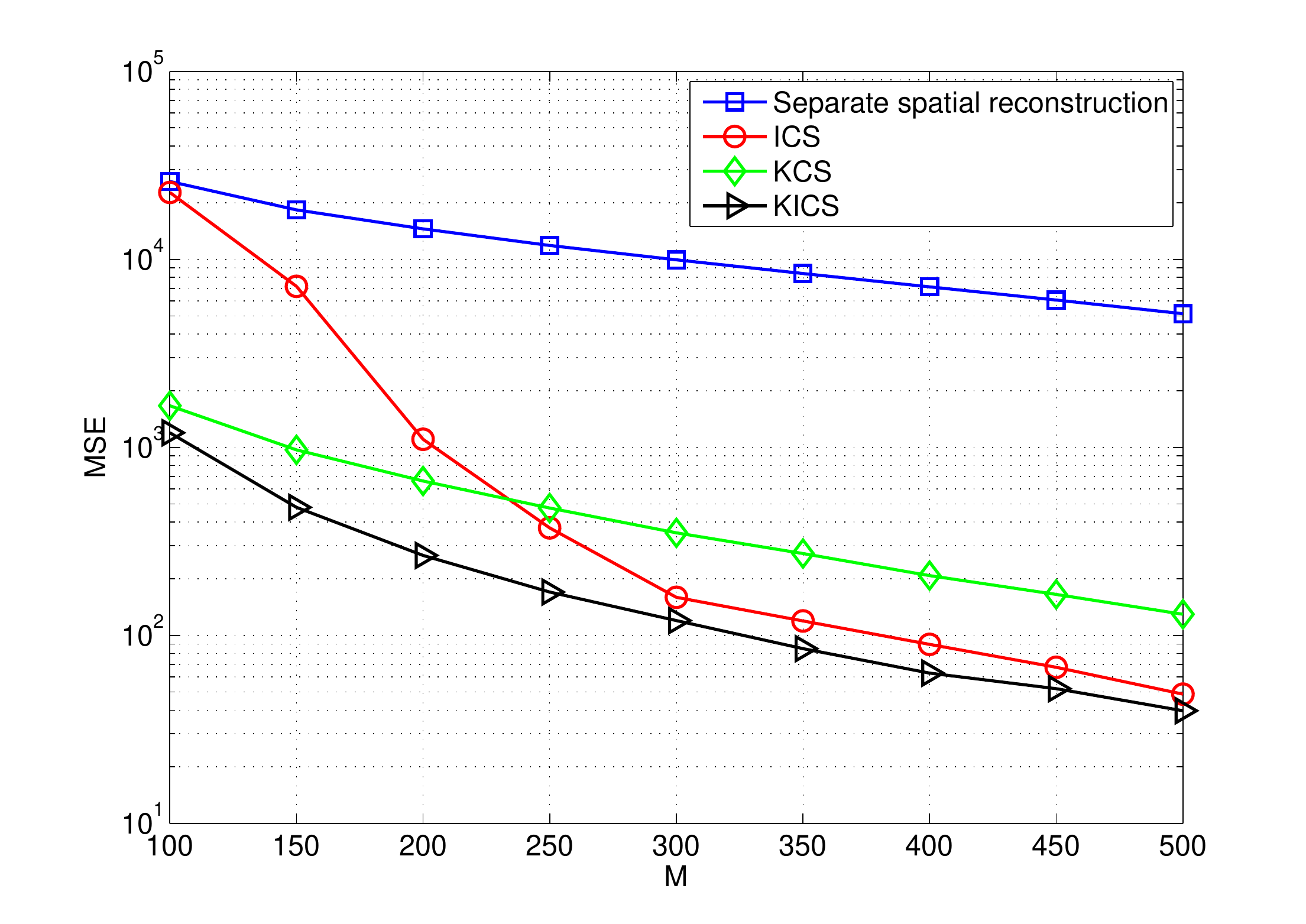}
\caption{AIRS: Reconstruction MSE vs. $M$ for different schemes.}
\label{fig:comparison_airs}
\end{center}
\end{figure}

Results in terms of MSE versus $M$ are shown in Figures \ref{fig:comparison_aviris}
 and \ref{fig:comparison_airs} for the AVIRIS and AIRS scenes, respectively. As can be
seen, S2D spatial reconstruction yields very large
mean-squared error (MSE), typically in excess of $5\cdot 10^4$ for AVIRIS and of $7\cdot 10^3$ for AIRS.
Considering that the average signal energy for this crop is equal to
$2.76\cdot 10^7$ for AVIRIS and $4.85\cdot 10^6$ for AIRS, spatial reconstruction yields an average
percentage error of nearly $\pm 4\%$ both test images,
which is inadequate for most
applications.

\begin{figure}
\begin{center}
 \includegraphics[width=0.47\textwidth]{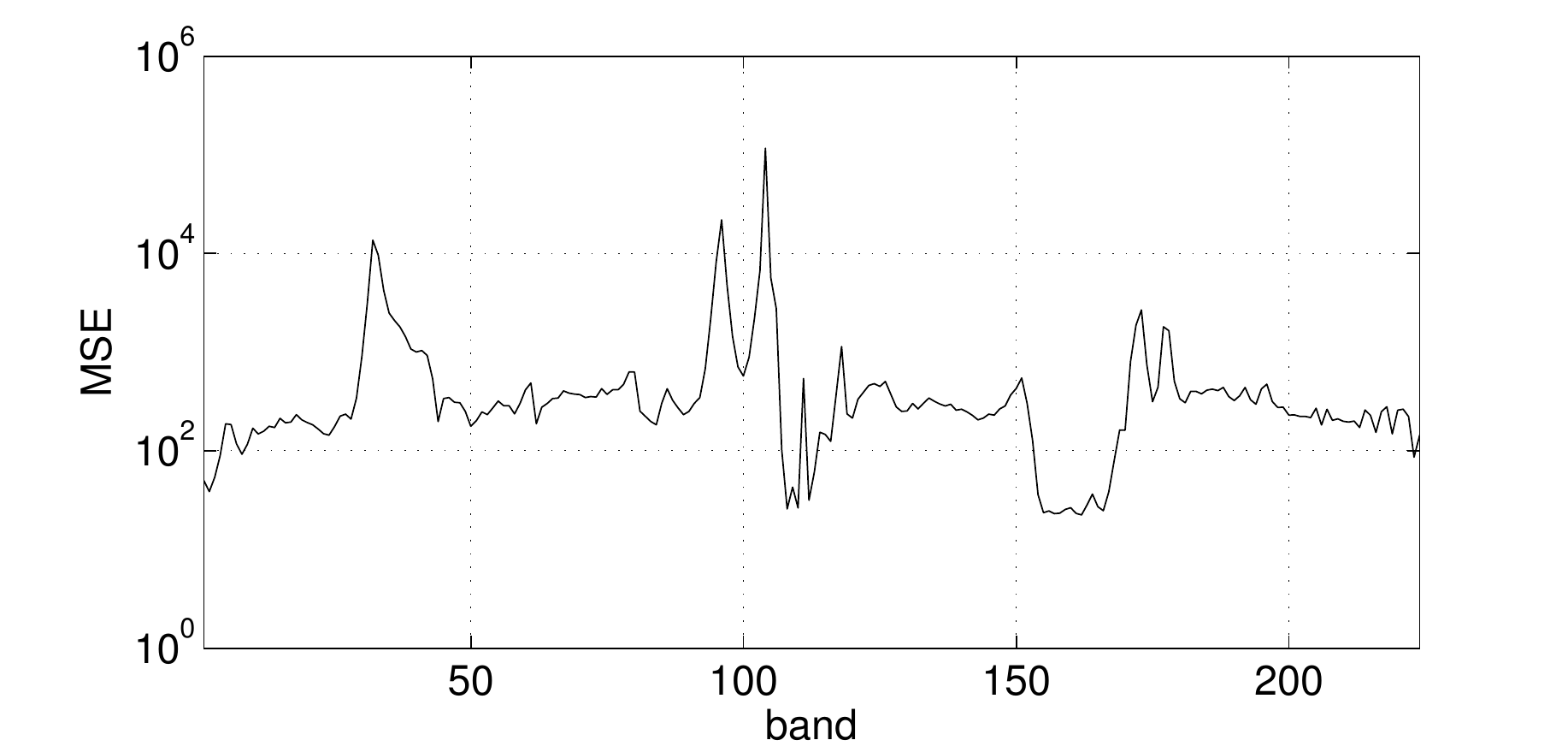} \\
\caption{AVIRIS: Reconstruction MSE for each band, $M = 450$.}
\label{fig:aviris_band}
\end{center}
\end{figure}

\begin{figure}
\begin{center}
 \includegraphics[width=0.47\textwidth]{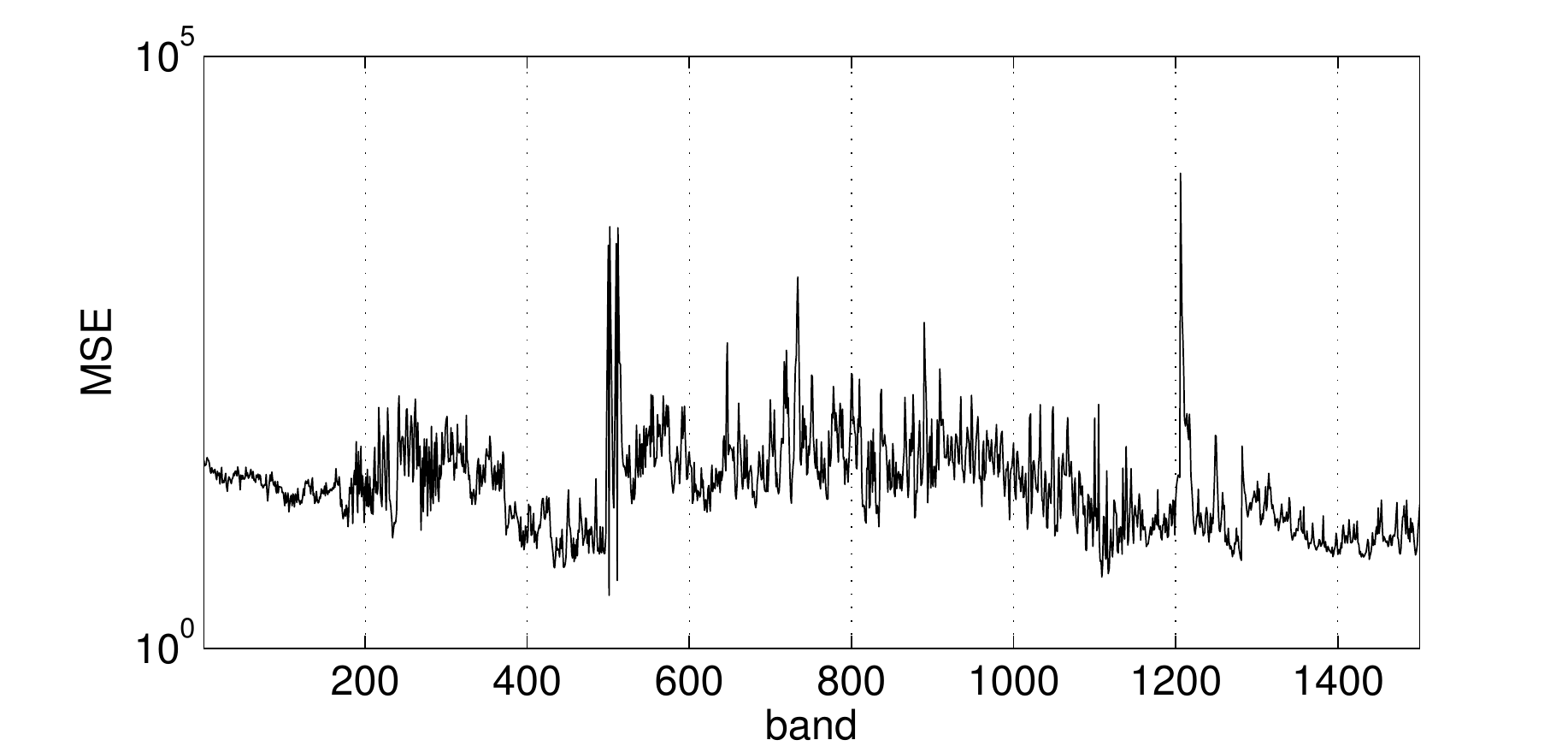}
\caption{AIRS: Reconstruction MSE for each band, $M = 450$.}
\label{fig:airs_band}
\end{center}
\end{figure}

As anticipated in Figure \ref{fig:aviris_2D_iterations}, the proposed ICS reconstruction algorithm allows to
improves the MSE up to a factor of 35 for high $M$, but it is not effective for low $M$. On the other hand,
 the 3D KCS reconstruction without iterative predictions performs quite well for low $M$ but its performance
 are not so good for high $M$, e.g., it is even worse than ICS for $M > 200-250$. Eventually, KICS gives
 the best performance over the whole range of considered $M$. In other words, combining 3D KCS with predictive
 CS allows to accurately reconstruct original images requiring a number of linear measurements much smaller than the original samples.
On the other hand, average results provide a somewhat biased picture though. In Fig.
\ref{fig:aviris_band} and \ref{fig:airs_band}, the individual MSE per band and for $M = 450$ obtained through KICS algorithm on AVIRIS and AIRS scene is shown, respectively. As
can be seen, in most bands the MSE is very small, between 100 and
400. The average MSE is biased by a relatively small number of bands
which are reconstructed with large error. Visual inspection shows
that e.g. band 104 is extremely noisy (hence not at all sparse) and
contains almost no information, while band 32 is misregistered with
respect to band 31, yielding poor prediction. This shows that, on
average, a much lower relative error can be achieved in
most bands, except for noisy bands, which are not very important
altogether, or misregistered bands, where improved prediction models
can be employed to improve the reconstruction.

\subsubsection{Separate acquisition of spectral rows}

\begin{figure}
\begin{center}
 \includegraphics[width=0.47\textwidth]{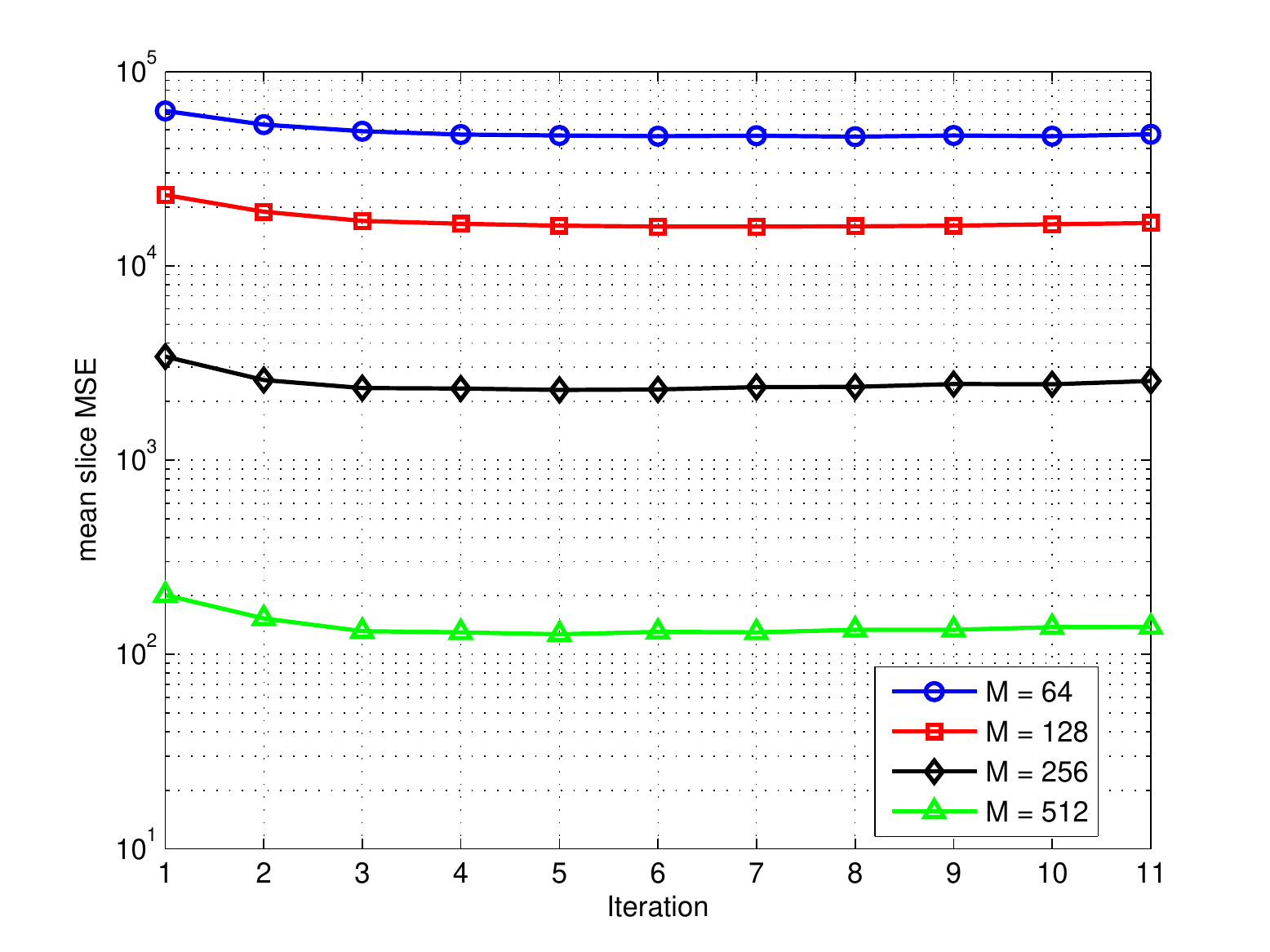}
\caption{MSE performance of ICS applied to spectral rows.}
\label{fig:airs9_rIter_cbMeas}
\end{center}
\end{figure}

Fig.~\ref{fig:airs9_rIter_cbMeas} refers to a $32\times 32\times 32$ crop of AIRS. Instead of taking separate measurements of spectral channels, here we separately acquire horizontal spectral rows $\F^i \in \Ri^{N_\mathsf{COL}\times N_\mathsf{BAND}}$ , where $\F^i = \mathcal{F}_{i,:,:}$ and $i = 1,\ldots,N_{\mathsf{ROWS}}$. The cube is then reconstructed using algorithm \ref{alg:3D_decoding_band_unique}, iterating over rows instead of wavelength. In this case, to predict spectral rows we use prediction filter \emph{P1} described in section \ref{sec:lp_2D}. Results show that separate reconstruction (the MSE of the initial step) applied to spectral rows leads to better performance than the same algorithm applied to spectral channels. Nevertheless, the iterative algorithm is less effective when iterating over rows than when iterating over wavelength. Results show that only a slight MSE gain can be obtained from iterative algorithm. This effect is due to stronger correlation along wavelength direction than between rows. Stronger correlation is better exploited by Compressed Sensing, leading to better separate reconstruction performance, while weaker correlation between rows yields only a minor contribution of the iterative algorithm.

\vspace{-2.5mm}
\section{Conclusions}\label{sec:conclusions}

In this paper, we proposed a general architecture for the acquisition and reconstruction of multidimensional correlated signals with manageable computational complexity. The acquisition is based on Compressed Sensing and consists in taking a sequence of separate random linear measurements of the signal, grouping subsets of the coordinates, in a progressive fashion. The reconstruction process implements an iterative architecture relying on linear prediction filters and the Compressed Sensing reconstruction of the prediction error, which is supposed to be more compressible than the original signal.
Then, we specialize this framework to 2D signals and 3D signals, proposing practical applications for these scenarios. For 2D signals, we envisage applications to \emph{flatbed scanners} and \emph{remote sensing}, which already perform progressive scanning. For 3D signals, the straightforward application is \emph{hyperspectral imaging}, even if applications related to \emph{video} could be also be imagined.
For both scenarios, we show that the performance in terms of MSE and speed of convergence depend on two factors. On one hand, the initial MSE of the algorithm depends on the initialization strategy. We show that the performance obtained by trivially initializing the algorithm with separate measurement reconstruction can be improved by using the so-called \emph{Kronecker Compressed Sensing}, which is able to capture the correlation in all dimensions at the cost of an increased computational complexity. On the other hand, the effectiveness of the iterative algorithm in terms of MSE gain strongly depends on the choice of the linear prediction filter and on the amount of signal correlation along the iteration dimension.

\ifCLASSOPTIONcaptionsoff
  \newpage
\fi

\end{document}